\documentclass[aps,prd,showpacs,superscriptaddress,preprintnumbers]{revtex4}
\usepackage{graphicx,float,wrapfig,subfigure}
\usepackage{amsfonts,amsmath,amssymb,amstext}
\usepackage{latexsym}
 \usepackage{bm}
\usepackage{color}
\usepackage[normalem]{ulem}

\newcommand{\be}{\begin{equation}}
\newcommand{\ee}{\end{equation}}
\newcommand{\ba}{\begin{eqnarray}}
\newcommand{\ea}{\end{eqnarray}}

\newcommand{\sign}{\,\mbox{sign}}

\definecolor{red}{rgb}{0.7,0,0}
\definecolor{green}{rgb}{0,0.5,0}

\begin{document}

\title{The pole and screening masses of neutral pion in hot and magnetized medium: a comprehensive study in the Nambu--Jona-Lasinio model}

\author{Bingkai Sheng}
\affiliation{ College of Physics, Jilin University, Changchun 130012, P.R. China}

\author{Yuanyuan Wang}
\affiliation{ College of Physics, Jilin University, Changchun 130012, P.R. China}

\author{Xinyang Wang}
\email{wangxy@ujs.edu.cn}
\affiliation{ Department of Physics, Jiangsu University, Zhenjiang 212013, P.R. China}

\author{Lang Yu}
\email{yulang@jlu.edu.cn}
\affiliation{ College of Physics, Jilin University, Changchun 130012, P.R. China}

\begin{abstract}
In this work, we investigate not only the pole masses but also the screening masses of neutral pions at finite temperature and magnetic field by utilizing the random phase approximation (RPA) approach in the
framework of the two-flavor Nambu--Jona-Lasinio (NJL) model. And two equivalent formalisms in the presence of a magnetic field, i.e. the Landau level representation (LLR) and the proper-time representation (PTR), are applied to obtain the corresponding analytical expressions of
the polarization functions (except the expressions for the pole masses in the PTR). In order to evaluate the applicable region of the low-momentum expansion (LME), we compare the numerical results within the full RPA (FRPA) with those within the reduced RPA (RRPA), i.e. the RPA in the LME. It is confirmed that the pole masses of $\pi^0$ in the FRPA suffer a sudden mass jump at the Mott transition temperature when in the presence of external magnetic field, and the Mott transition temperature is catalyzed by the magnetic field. And by analyzing the behaviors of the directional sound velocities of $\pi^0$, which are associated with the breaking of the Lorentz invariance by the heat bath and the magnetic field, we clarify the two problems existing in previous literatures: one is that the transverse sound velocities in the medium are always larger than unity and thus violate the law of causality on account of the non-covariant regularization scheme, the other is that the longitudinal sound velocities are identically equal unity at finite temperature on account of the limitation of the derivative expansion method used.
\end{abstract}

\maketitle

\section{Introduction}

Extremely strong magnetic fields are expected to exist in several important high-energy physical systems, such as the early universe~\cite{Vachaspati:1991nm,Enqvist:1993np}, compact stars~\cite{Duncan:1992hi}, and the noncentral heavy ion collisions~\cite{Skokov:2009qp,Voronyuk:2011jd,Bzdak:2011yy,Deng:2012pc}. As a consequence, in the last few decades, more and more attention have been paid to the influence of strong magnetic fields on the strongly interacting matter. Theoretically, it has been known that, the interplay between non-perturbative properties of Quantum chromodynamics (QCD) and the strong magnetic field, might give rise to a variety of intriguing phenomena, for example, the chiral magnetic effect (CME)~\cite{Kharzeev:2007tn,Kharzeev:2007jp,Fukushima:2008xe}, magnetic catalysis~\cite{Klevansky:1989vi,Klimenko:1990rh,Gusynin:1995nb,Shovkovy:2012zn} and inverse magnetic catalysis~\cite{Bali:20111213}, vacuum superconductivity~\cite{Chernodub:2010qx,Chernodub:2011mc}, and so on. In particular, some of these phenomena are associated with the hadron properties in external magnetic fields. For one thing, in order to confirm or exclude the existence of the charge rho meson condensation in strong magnetic fields, i.e. the electromagnetic superconductivity of vacuum, the behaviors of rho meson masses dependent on the magnetic field strength have been computed by various effective theories and models~\cite{Chernodub:2010qx,Chernodub:2011mc,Callebaut:2011uc,Ammon:2011je,
Cai:2013pda,Frasca:2013kka,Andreichikov:2013zba,Wang:phd,Liu:2014uwa,
Liu:2015pna,Liu:2016vuw,Kawaguchi:2015gpt,
Zhang:2016qrl,Ghosh:2016evc,Ghosh:2017rjo} as well as lattice QCD simulations~\cite{Hidaka:2012mz,Luschevskaya:2014mna,Luschevskaya:2015bea,
Bali:2017ian,Ding:2020jui}. For another, since neutral pions are Nambu-Goldstone bosons of the chiral symmetry breaking, the modifications of their properties in an
external magnetic field will help to understand the effects of magnetic fields on the chiral phase transition, and thus are extensively investigated recently~\cite{Klevansky:1991ey,Andersen:2012zc,Fayazbakhsh:2012vr,Fayazbakhsh:2013cha,
Orlovsky:2013wjd,Luschevskaya:2014lga,Avancini:2015ady,Simonov:2015xta,Luschevskaya:2015cko,
Bali:2015vua,Avancini:2016fgq,Hattori:2015aki,Mao:2017wmq,GomezDumm:2017jij,Aguirre:2017dht,
Wang:2017vtn,Liu:2018zag,Ayala:2018zat,Avancini:2018svs,Chaudhuri:2019lbw,Coppola:2019uyr,
Das:2019ehv,Ding:2020hxw}. Furthermore, there are some other works involving heavy mesons~\cite{Marasinghe:2011bt,Machado:2013rta, Alford:2013jva,Machado:2013yaa,
Cho:2014exa,Cho:2014loa,Dudal:2014jfa,Bonati:2015dka,Gubler:2015qok,
Yoshida:2016xgm,Reddy:2017pqp,CS:2018mag} and baryons~\cite{Tiburzi:2008ma,
Andreichikov:2013pga,Tiburzi:2014zva,Haber:2014zba,He:2016oqk,Deshmukh:2017ciw,
Yakhshiev:2019gvb} in the magnetic field.

In the present paper, we will focus on studying the neutral pion masses, including not only pole masses but also screening masses, at finite temperature and magnetic field via RPA approach in the two-flavor NJL model. As mentioned in the last paragraph, there are a lot of works that have studied the pole masses of $\pi^0$ mesons under a constant external magnetic field, but only several of them explored the screening masses of them correspondingly~\cite{Fayazbakhsh:2012vr,Fayazbakhsh:2013cha,Wang:2017vtn}. On the other hand, in Refs.~\cite{Fayazbakhsh:2012vr,Fayazbakhsh:2013cha,Wang:2017vtn},
the pion masses, including the pole and screening masses, were obtained by employing the derivative expansion method in the NJL model, which is just equivalent to the RPA in the LME (or call it the reduced RPA)~\cite{Wang:2017vtn}. Therefore, we will complete the analytical derivations and numerical calculations for both the pole and screening masses of neutral pions within the FRPA under the magnetic field. Furthermore, due to the explicit breaking of the Lorentz invariance by the magnetic field and the temperature bath, there is not only an anisotropy between the transverse and longitudinal directions with respect to the direction of the external magnetic field, but also another one between the temporal and spatial directions. Both of them will be thoughtfully discussed in this paper, whereas the latter one was neglected in Refs.~\cite{Fayazbakhsh:2012vr,Fayazbakhsh:2013cha,Wang:2017vtn} as a result of the defects of the derivative expansion method at finite temperature.

Based on the strategy in Refs.~\cite{Fayazbakhsh:2012vr,Fayazbakhsh:2013cha}, a
nontrivial anisotropic energy dispersion relation for neutral pions in the hot and magnetized medium (the magnetic field is assumed to be in the positive $z$ direction without loss of generality) is introduced by
\be\label{EDR}
E^2=u_{\perp}^2 \mathbf{q}_{\perp}^2+u_{\parallel}^2q_3^2+m_{\pi_0,pole}^2,
\ee
where $u_{\perp}=u_1=u_2$ and $u_{\parallel}=u_3$ represent the pion transverse and longitudinal velocities ($u_i$ is the pion sound velocity in the $q_i$ direction), respectively and $m_{\pi_0,pole}$ is the pole mass of neutral pions. Here, according to the patterns of Lorentz symmetry breaking by the temperature bath and the magnetic field, $u_{\perp}=u_{\perp}(B,T)$ is a quantity dependent on both $T$ and $B$, while $u_{\parallel}=u_{\parallel}(T)$ is $T$-dependent only. Correspondingly, the transverse and longitudinal screening masses of the neutral pion are defined by $m_{\pi_0,scr,\perp}=\frac{m_{\pi_0,pole}}{u_{\perp}}$ and $m_{\pi_0,scr,\parallel}=\frac{m_{\pi_0,pole}}{u_{\parallel}}$. Please note that in Refs.~\cite{Fayazbakhsh:2012vr,Fayazbakhsh:2013cha}, $u_{\perp}$ and $u_{\parallel}$ were defined as the refraction indices mistakenly. Thus, in our paper, the definitions of the directional refraction
indices are given by $n_{\perp}=1/u_{\perp}$ and $n_{\parallel}=1/u_{\parallel}$.

Now, we begin to consider three special cases for Eq.~(\ref{EDR}) in the following: First, at $T = 0$ and $B=0$, we should have $u_{\perp}=u_{\parallel}=1$, and thus $m_{\pi_0,pole}=m_{\pi_0,scr,\perp}=m_{\pi_0,scr,\parallel}$ because of the Lorentz invariance. It implies that, in the chiral limit, massless pions propagate at the speed of light in vacuum. Second, at $T\neq 0$ but $B=0$, since the temperature breaks the boosts of Lorentz symmetry, it is expected that
$u_{\perp}=u_{\parallel}=u\neq 1$ and $m_{\pi_0,pole}\neq m_{\pi_0,scr,\perp}=m_{\pi_0,scr,\parallel}$. And as discussed in Refs.~\cite{Pisarski:1996mt,Pisarski:1996yc,Pisarski:1996zv}, the sound velocity $u$ is always smaller than unity (i.e. $u<1$), which means massless pions, in the chiral limit, travel at a speed
slower than the speed of light in the hot medium, obeying the law of causality. Third, for $B\neq0$ but $T = 0$, the anisotropy arised by the constant magnetic field leads to $u_{\perp}\neq u_{\parallel}$ but $u_{\parallel}=1$, as well as $m_{\pi_0,scr,\perp}\neq m_{\pi_0,scr,\parallel}$ and $m_{\pi_0,pole}=m_{\pi_0,scr,\parallel}$. In order to satisfy the causality, we expect that $u_{\perp}<1$ and $m_{\pi_0,pole}=m_{\pi_0,scr,\parallel}<m_{\pi_0,scr,\perp}$. It is found that, our statements are consistent with the results in Ref.~\cite{Wang:2017vtn}, but in contrast with those in Refs.~\cite{Fayazbakhsh:2012vr,Fayazbakhsh:2013cha}, although all calculations were done by the same derivative expansion method in the NJL model. According to argument in Ref.~\cite{Mao:2017wmq},
this problem is caused by the regularization schemes used in these two papers: the former one uses the covariant regularization scheme (the Pauli-Villars scheme), while the latter one uses the smooth noncovariant
cutoff scheme. This is the reason why we choose the Pauli-Villars scheme
in this paper. Finally, to summarize the above discussions, when $T\neq 0$ and $B\neq0$, we arrive at the conclusions: $u_{\perp}< u_{\parallel}<1$ and
$m_{\pi_0,pole}<m_{\pi_0,scr,\parallel}<m_{\pi_0,scr,\perp}$. However, in Refs.~\cite{Fayazbakhsh:2012vr,Fayazbakhsh:2013cha,Wang:2017vtn}, it is claimed that the longitudinal velocity $u_{\parallel}$ of neutral pions is equal to unity always and independent of $T$. The source of the error is because the derivative expansion method ignores the breaking of Lorentz invariance by the heat bath: more explicitly, in the one-quark-loop polarization function calculations, the static limit ($q_0=0$, $q=|\mathbf{q}|\rightarrow 0$) do not commute with the long-wavelength limit ($q_0\rightarrow 0$, $q=|\mathbf{q}|=0$) at finite temperature\cite{Weldon:1992bv}, and these two limits solely commute at $T=0$. In addition, the erroneous results $u_{\perp}>u_{\parallel}$ and $m_{\pi_0,scr,\perp}<m_{\pi_0,scr,\parallel}$ at nonzero $T$ and $B$ were also obtained in Ref.~\cite{Fayazbakhsh:2012vr,Fayazbakhsh:2013cha} due to the non-covariant regularization scheme, which will be corrected by employing the Pauli-Villars scheme in this paper.

The work is organized as follows. In Sec.~II, we will first introduce the two-flavor NJL model in the presence of an
external magnetic field and derive the gap equation for the quark mass in the mean field approximation at finite temperature and magnetic field. And then, the pole and screening masses of neutral pions in the hot and magnetized medium are calculated based on the standard RPA approach, where we make use of two equivalent formalisms in the magnetic field, i.e. the LLR and the PTR. And notice that, since it is difficult to compute the pole masses at finite temperature in terms of the PTR, we only deal with them in the LLR. Next, in Sec.~III, we will show our numerical results not only within the FRPA but also within the RRPA. Finally, the summary and conclusions will be presented in Sec.~IV.

\section{formalism}

\subsection{NJL Model and the Gap Equation}

The Lagrangian density of two-flavor NJL model ~\cite{Nambu:1961tp,Nambu:1961fr} under a constant external magnetic field is given by
\ba
{\cal{L}}&=&\bar{\psi}(i\not{\!\!D}-\hat{m})\psi+G\left[(\bar{\psi}\psi)^2
   +(\bar{\psi}i\gamma^5\vec{\tau}\psi)^2\right].
\label{eq:L:basic}
\ea
Where $\psi=\left(\begin{array}{c}
                    u \\
                    d
                   \end{array}
            \right)$
represents the quark fields of two light flavors, $\vec{\tau}=(\tau^1,\tau^2,\tau^3)$ is the isospin Pauli matrix and $G$ is the coupling constant corresponding to the (pseudo)scalar channel. The current mass matrix $\hat{m}=\text{diag}(m_u,m_d)$, and we assume that $m_u=m_d=m_0$. As for the covariant derivative, $D_{\mu}=\partial_{\mu}+i \hat{Q} e A_{\mu}^{ext}$, couples quarks to an external magnetic field $\bm{B}=(0,0,B)$ along the positive $z$ direction via a background field, for example, $A_{\mu}^{ext}=(0,0,-Bx,0)$. Besides, $\hat{Q}=\mathrm{diag}(Q_u,Q_d)=\mathrm{diag}(2/3,-1/3)$ is  a diagonal quark charge matrix in the flavor space, and $e$ is the absolute value.

In the mean field approximation, the constituent quark mass $m$ is determined by the gap equation~\cite{Florkowski:1997pi,Klevansky:1992qe}
\be \label{eq:gapequationmq}
m=m_0-2G \left<\bar{\psi}\psi\right>,
\ee
which is obtained by minimizing the thermodynamical potential. And $\left<\bar{\psi}\psi\right>=-\mathrm{Tr} S(u,u)$ with the quark propagator in the mean field approximation $S=i(i\not{\!\!D}-\hat{m})^{-1}=\mathrm{diag}(S_u,S_d)$. In the presence of a constant magnetic field, the quark propagator $S(u,u^\prime)$ can be expressed in the LLR~\cite{Miransky:2015ava},
and it takes the following form:
\begin{eqnarray}\label{quark-GB}
S(u,u^\prime) &=& e^{i\Phi(\mathbf{r}_\perp,\mathbf{r}_\perp^\prime)}\widetilde{S}(u-u^\prime), \\
\widetilde{S}(u-u^\prime)&=& \int \frac{d^4p}{(2\pi)^4}
e^{-ip\cdot(u-u^\prime)}
\widetilde{S}(p),
\end{eqnarray}
where $u=(t, x, y, z)$, $\mathbf{r}_\perp=(x,y)$ and $\Phi(\mathbf{r}_\perp,\mathbf{r}_\perp^\prime)=-Q_feB(x+x^{\prime})(y-y^{\prime})/2$
is the Schwinger phase \cite{Schwinger} for the vector potential in the above Landau gauge. In momentum space, the Fourier transform $\widetilde{S}(p)$ of the translationally invariant part $\widetilde{S}(u-u^\prime)$ is given by
\begin{equation}
\widetilde{S}(p) = ie^{-p_{\perp}^2 l^{2}}\sum_{n=0}^{\infty}
\frac{(-1)^nD_{n}(p)}{p_0^2-p_{3}^2-m^2-2n|Q_feB|},
\label{GDn-new}
\end{equation}
where the numerator of the $n$th Landau level contribution is determined by
\begin{equation}
D_{n}(p) =
2\left[p^0\gamma^0 -p^{3}\gamma^3 + m \right]
\left[{\cal P}_{+}L_n\left(2 p_{\perp}^2 l^{2}\right)
-{\cal P}_{-}L_{n-1}\left(2 p_{\perp}^2 l^{2}\right)\right]
 + 4(p^1\gamma^1+p^2\gamma^2)
 L_{n-1}^1\left(2 p_{\perp}^2 l^{2}\right).
\label{Dn}
\end{equation}
Here $L_{n}^\alpha(z)$ are the generalized Laguerre polynomials,
${\cal P}_{\pm}\equiv \frac12 \left(1\pm i s_\perp \gamma^1\gamma^2\right)$
are spin projectors, and $l=\sqrt{1/|Q_feB|}$ is the magnetic length. By definition,
$s_\perp=\sign (Q_feB)$ and $L_{-1}^\alpha \equiv 0$.

By making use of
Eqs.~(\ref{quark-GB})-(\ref{Dn}) and performing Matsubara frequency summation, the gap equation of Eq.~(\ref{eq:gapequationmq}) at finite temperature and magnetic field can be explicitly written as follows,
\be \label{eq:gapequationmq:vac:eB}
m=m_0+4G N_c m I_{1}(m^2),
\ee
and the function $I_{1}(m^2)$ is defined as
\be \label{I1}
I_{1}(m^2)=\sum_{f=u,d}\frac{|Q_f eB|}{2\pi} \sum_{n=0}^{\infty} \beta_n I'_{1}(m_{n,f}^2),
\ee
where $\beta_n=2-\delta_{n,0}$, and
\be \label{eq:I1:vac:eB}
I'_{1}(m_{n,f}^2)=-T\sum^{\infty}_{l=-\infty}\int\frac{dp_3}{2\pi}\frac{1}{(i\omega_l)^2-p_3^2-m_{n,f}^2+i\epsilon}
=\int\frac{dp_3}{2\pi}\frac{1}{2E_{n,f}}\frac{e^{E_{n,f}/T}-1}{e^{E_{n,f}/T}+1},
\ee
with $i \omega_{l}=i(2 l+1) \pi T$, $m_{n,f}^2=2n|Q_f eB|+m^2$ and  $E_{n,f}=\sqrt{p_3^2+m_{n,f}^2}$. And $I'_{1}(m_{n,f}^2)$ can be divided into a sum of two contributions--the vacuum part and the temperature part,
\be
I'_{1}(m_{n,f}^2) = I'_{1,vac}(m_{n,f}^2)+I'_{1,tem}(m_{n,f}^2),
\ee
with
\ba
I'_{1,vac}(m_{n,f}^2) &=& \int\frac{dp_3}{2\pi}\frac{1}{2E_{n,f}}, \\
I'_{1,tem}(m_{n,f}^2) &=&
\int\frac{dp_3}{2\pi}\frac{1}{2E_{n,f}}\frac{(-2)}{e^{E_{n,f}/T}+1}.
\ea
Notice that the vacuum part is divergent, while the temperature part is convergent and vanishes at $T=0$.

Since the NJL model is not renormalizable, regularization is needed in the model calculations. In this paper, we shall use the Pauli-Villars
regularization scheme, which preserves the Lorentz symmetry. And although the 3-dimensional cutoff regularization scheme is commonly used,
it breaks the Lorentz invariance and leads to an unphysical oscillation~\cite{Florkowski:1997pi}. In the Pauli-Villars scheme, an integral $I(m)$ is regularized as
\be \label{eq:PV}
I(m)=\sum_{\alpha=0}^N C_{\alpha}I(m_{\alpha}),
\ee
where $m_{\alpha}=\sqrt{m^2+a_{\alpha}\Lambda^2}$ are masses of auxiliary particles. The parameters $a_{\alpha}$ and $C_{\alpha}$ are determined
by the conditions $a_0=0$, $C_0=1$ and $\sum_{\alpha=0}^N C_{\alpha} m_{\alpha}^{2L}=0$ for $L=0,1,...N-1$. In this article, we make the choice
$(C_0,C_1,C_2)=(1,1,-2)$ and $(a_0,a_1,a_2)=(0,2,1)$ . And then the parameters $G\Lambda^2=2.87$, $\Lambda=851$ MeV, and
$m_0=5.2$ MeV are fixed by the decay constant $f_{\pi}=93$ MeV, $\left<\bar{u}u\right>=(-250\, \text{MeV})^3$, and $m_{\pi}=135$ MeV as in Ref.~\cite{Klevansky:1992qe}.

Therefore, in the Pauli-Villars scheme,
the corresponding expressions are given by
\be \label{eq:I1PV:vac:eB}
I'^{PV}_{1,vac}(m_{n,f}^2) = -\frac{1}{4\pi}\sum_{\alpha=0}^NC_{\alpha}\ln(m^2_{n,f,\alpha}),
\ee
and
\be \label{eq:I1PV:tem:eB}
I'^{PV}_{1,tem}(m_{n,f}^2) = \sum_{\alpha=0}^N C_{\alpha} I'_{1,tem}(m_{n,f,\alpha}^2),
\ee
where $m^2_{n,f,\alpha}=m_{n,f}^2+a_{\alpha}\Lambda^2$. Now, $I'^{PV}_{1}(m_{n,f}^2)=I'^{PV}_{1,vac}(m_{n,f}^2)+I'^{PV}_{1,tem}(m_{n,f}^2)$
can be used to calculate the constituent quark mass $m$ at finite $T$ and $eB$ by the gap equation.

Alternatively, the translationally invariant part $\widetilde{S}(u-u')$ can also be expressed in the mixed coordinate-momentum space~\cite{Miransky:2015ava}
\begin{eqnarray}\label{quark-mixed-GB1}
\widetilde{S}(u-u')&=& \int \frac{dp_0dp_3}{(2\pi)^2}
e^{-ip_0t+ip_3z}
\widetilde{S}(p_0, p_3 ;\mathbf{r}_\perp-\mathbf{r}_\perp'),
\\
\widetilde{S}(p_0, p_3 ;\mathbf{r}_\perp-\mathbf{r}_\perp') &=& i\frac{e^{-\mathbf{r}_\perp^2/(4l^2)}}{2\pi l^2}
\sum_{n=0}^{\infty}
\frac{\widetilde{D}_{n}(p_0,p_3 ;\mathbf{r}_\perp-\mathbf{r}_\perp')}{p_0^2-p_{3}^2-m^2-2n|Q_feB|},
\label{quark-mixed-GB2}
\end{eqnarray}
where the numerator of the $n$th Landau level contribution is determined by
\begin{equation}
\widetilde{D}_{n}(p_0,p_3 ;\mathbf{r}_\perp-\mathbf{r}_\perp') = \left[p^0\gamma^0 -p^{3}\gamma^3 + m \right]
\left[{\cal P}_{+}L_n\left(\xi\right)
+{\cal P}_{-}L_{n-1}\left(\xi\right)\right]
-\frac{i}{l^2}(\mathbf{r}_{\perp}-\mathbf{r}_{\perp}')\cdot\bm{\gamma}_{\perp}
 L_{n-1}^1\left(\xi\right)
\label{Dn-alt}
\end{equation}
with $\xi=\frac{(\mathbf{r}_{\perp}-\mathbf{r}_{\perp}')^2}{2l^{2}}$ and $\bm{\gamma}_{\perp}=(\gamma^1,\gamma^2)$.

Further more,  the sum over Landau levels can be easily performed in Eq.~(\ref{GDn-new}) with the help of the summation formula for Laguerre polynomials~\cite{Gradstein},
\begin{equation}
\sum_{n=0}^{\infty} L_{n}^{\alpha}(x) z^{n}=(1-z)^{-(\alpha+1)} \exp \left(\frac{x z}{z-1}\right).
\end{equation}
Then we obtain the propagator $\widetilde{S}(p) $ in the proper-time representation~\cite{Schwinger},
\begin{eqnarray}\label{quark-pt-GB}
\tilde{S}(p)&=& \int_{0}^{\infty} d s e^{i s\left(p_0^{2}-p_3^{2}-m^{2}-p_{\perp}^{2} \frac{\tan (Q_feBs)}{Q_feBs}\right)}\left[\gamma^{\mu} p_{\mu}+m+\left(p^{1} \gamma^{2}-p^{2} \gamma^{1}\right) \tan (Q_feBs) \right]\left[1-\gamma^{1} \gamma^{2} \tan (Q_feBs)\right].
\end{eqnarray}
By utilizing Eq.~(\ref{quark-pt-GB}), the functions $I_{1,vac}(m^2)$ for $T=0 $ and $I_{1}(m^2)$ for $T\neq 0$ can be rewritten in the following forms respectively,
\begin{equation}\label{I1-vac-PT}
I_{1,vac}(m^2)=\frac{1}{8\pi^2}\sum_{f=u,d}Q_{f} e B \int_{0}^{\infty} ds \frac{e^{-m^{2} s}}{s}\coth\left(Q_{f} e B s\right) ,
\end{equation}
and
\ba\label{I1-T-PT}
I_{1}(m^2)
&=&\frac{T}{4\pi^{\frac{3}{2}}}\sum_{f=u,d}Q_{f} e B \int_{0}^{\infty} ds   \frac{e^{-m^{2} s}}{\sqrt{s}}\coth\left(Q_{f} e B s\right) \left(\sum_{l=-\infty}^{\infty} e^{-s\omega_l^2}\right) \nonumber\\
&=&\frac{T}{4\pi^{\frac{3}{2}}}\sum_{f=u,d}Q_{f} e B \int_{0}^{\infty} ds \frac{e^{-m^{2} s}}{\sqrt{s}}\coth\left(Q_{f} e B s\right) \theta_{2}\left(0, e^{-4 \pi^{2}s T^{2}}\right),
\ea
where the Jacobi theta function $\theta_{2}(0, q)=2 \sqrt[4]{q} \sum_{n=0}^{\infty} q^{n(n+1)}$. We follow the Pauli-Villars regularization procedure above, so that Eqs.~(\ref{I1-vac-PT}) and (\ref{I1-T-PT}) can be written as
\ba\label{I1-vac-PT-PV}
I_{1,vac}^{PV}(m^2)
&=&\Delta I_{1,vac}^{PV}(m^2)+I_{1,vac}^{PV}(m^2)|_{B=0} \nonumber \\
&=&\frac{1}{8\pi^2}\sum_{\alpha=0}^N C_{\alpha}\sum_{f=u,d}Q_{f} e B \int_{0}^{\infty} ds \frac{e^{-m_{\alpha}^2 s}}{s}\Big[\coth\left(Q_{f} e B s\right)-\frac{1}{Q_{f} e B s}\Big]+ I_{1,vac}^{PV}(m^2)|_{B=0},
\ea
and
\ba\label{I1-T-PT-PV}
I_{1}^{PV}(m^2)&=&\Delta I_{1}^{PV}(m^2)+I_{1}^{PV}(m^2)|_{B=0} \nonumber \\
&=&\frac{T}{4\pi^{\frac{3}{2}}}\sum_{\alpha=0}^N C_{\alpha}\sum_{f=u,d}Q_{f} e B \int_{0}^{\infty} ds \frac{e^{-m_{\alpha}^{2} s}}{\sqrt{s}}\Big[\coth\left(Q_{f} e B s\right)-\frac{1}{Q_{f} e B s}\Big] \theta_{2}\left(0, e^{-4 \pi^{2}s T^{2}}\right)\nonumber \\&&+I_{1}^{PV}(m^2)|_{B=0},
\ea
where
\be\label{I1-PV-B0}
I_{1}^{PV}(m^2)|_{B=0}=I_{1,vac}^{PV}(m^2)|_{B=0}+I_{1,tem}^{PV}(m^2)|_{B=0},
\ee
with
\be\label{I1-PV-vac-B0}
I_{1,vac}^{PV}(m^2)|_{B=0} =\frac{N_f}{8\pi^2}\sum_{\alpha=0}^N C_{\alpha}m_{\alpha}^2\ln(m_{\alpha}^2)
\ee
and
\be\label{I1-PV-tem-B0}
I_{1,tem}^{PV}(m^2)|_{B=0} =\frac{N_f}{2\pi^2}\sum_{\alpha=0}^N C_{\alpha}\int_0^{\infty} dp\frac{ p^2}{\sqrt{p^2+m_{\alpha}^2}}\frac{(-2)}{e^{\sqrt{p^2+m_{\alpha}^2}/T}+1},
\ee
which is obtained by using the quark propagator in vacuum $S(p)=\frac{i}{p \!\!\!/-m}$.

It will be shown later that, compared with the quark propagator in the LLR in the momentum space, it is more convenient to derive the analytical equations for meson transverse screening masses by using the quark propagator either in the LLR in the mixed coordinate-momentum space or in the PTR. Hence, in the following subsections, we will first derive the equations for the pole masses and longitudinal screening masses of neutral mesons by using Eq.~(\ref{GDn-new}), and then show the derivations of the equations for the transverse screening masses by using Eqs.~(\ref{quark-mixed-GB2}) and (\ref{quark-pt-GB}), respectively.

\subsection{meson pole and screening masses at finite T and eB}

In the framework of the NJL model, the quark-antiquark $T$ matrix for the meson channel $M$ is constructed by using the random phase approximation(RPA),
\be \label{eq:RPA}
T_M=\frac{2G}{1-2G\Pi_M},\,\,\,
\ee
and we focus on $M=\pi^0$ in this paper. And in momentum space, by introducing the propagator of Eq.~(\ref{GDn-new}), the meson polarization function $\Pi_{\pi^0}$ in the magnetic field at $T=0$ takes the form:
 \be\label{pionpolar}
 \Pi_{\pi^0,vac}(q )=-i\int\frac{d^4p}{(2\pi)^4}\text{Tr}[i\gamma_5\tau^3\widetilde{S}(p)i\gamma_5\tau^3\widetilde{S}(p-q)].
 \ee
Note that two Schwinger phases of the quark-antiquark pair cancel with each other in the neutral meson polarization function. At finite $T$, the
corresponding expression for $\Pi_{\pi^0}$ is obtained by the replacement
\ba
p_{0} &\rightarrow& i \omega_{l}=i(2 l+1) \pi T ,\\
\int \frac{d^{4} p}{(2 \pi)^{4}} &\rightarrow& i T \sum_{l=-\infty}^{\infty} \int \frac{d^{3} p}{(2 \pi)^{3}}.
\ea

Since $T_M$ matrix is interpreted as an effective meson propagator, the pole mass $m_{pole}$ (setting $q_1=q_2=q_3=0$) and the screening masses $m_{scr,i}$ in $q^i$ direction (setting $q_0=0$, and $q_j=0$ for $j\neq i$) can be solved by following equations, respectively,
\be\label{Gappole}
1-2G \Pi_{\pi^0}(q_0^2=m^2_{pole},\,0)=0
\ee
and
\be\label{Gapscr}
1-2G \Pi_{\pi^0}(0,\,q_i^2=-m^2_{scr,i})=0.
\ee
When $m_{pole}$ ($m_{scr}$) exceeds two times quark
mass, we need to make a replacement $m_{pole} \rightarrow m_{pole}-i \frac{\Gamma_{pole}}{2}$ ($m_{scr} \rightarrow m_{scr}-i \frac{\Gamma_{scr}}{2}$), and the mass $m_{pole}$ ($m_{scr}$) and its width $\Gamma_{pole}$ ($\Gamma_{scr}$) are determined by the corresponding complex equation. For simplicity, in this article we implicitly neglect the widths of the mesons and define the masses by the real parts of Eqs.~(\ref{Gappole}) and~(\ref{Gapscr}):
\be\label{Gappole-re}
1-2G \,\text{Re}[\Pi_{\pi^0}(q_0^2=m^2_{pole},\,0)]=0
\ee
and
\be\label{Gapscr-Re}
1-2G\, \text{Re}[ \Pi_{\pi^0}(0,\,q_i^2=-m^2_{scr,i})]=0.
\ee

Now, we begin with the derivations of the equations for the pole masses. By working in the rest frame of mesons, i.e. $q^{\mu}=(q^0,0,0,0)$, and using the orthogonal relationship for the Laguerre polynomials, the polarization function for $\pi^0$ pole masses at finite $T$ and $eB$ can be written as
\be\label{eq:PI:pi0:pole:eB}
\Pi_{\pi^0}(q_0^2,0)=N_c\left[2I_1(m^2)-q_0^2I_2(m^2,q_0^2,0)\right],
\ee
where
\be \label{eq:I2B:vac}
I_{2}(m^2,q_0^2,0)=\sum_{f=u,d}\frac{|Q_f eB|}{2\pi} \sum_{n=0}^{\infty} \beta_n I'_{2}(m^2,q_0^2,0),
\ee
and $I'_{2}(m^2,q_0^2,0)$ is obtained by taking analytical continuation $i\omega_j=i 2\pi j T \rightarrow q^0$ for
$I'_{2}(m^2,i\omega_j,0)$, with
\be \label{eq:I2PVeB}
I'_{2}(m^2,i\omega_j,0)=-T\sum^{\infty}_{l=-\infty}\int\frac{dp_3}{2\pi}
\frac{1}{(i\omega_l)^2-E_{n,f}^2}\frac{1}{(i\omega_l-i\omega_j)^2-E_{n,f}^2}.
\ee
Besides, it is shown that $I'_{2}(m^2,q_0^2,0)$ can also be separated into two terms, i.e. the vacuum and temperature parts,
\be \label{eq:I2PVeB}
I'_{2}(m^2,q_0^2,0)
=I'_{2,vac}(m^2,q_0^2,0)+I'_{2,tem}(m^2,q_0^2,0),
\ee
where
\be \label{eq:I2PV:vac:eB}
I'_{2,vac}(m^2,q_0^2,0)
=-\int\frac{dp_3}{2\pi}\frac{1}{4E_{n,f}(E_{n,f}^2-\frac{q_0^2}{4})},
\ee
and
\be \label{eq:I2PV:tem:eB}
I'_{2,tem}(m^2,q_0^2,0)=-\int\frac{dp_3}{2\pi}
\frac{1}{4E_{n,f}(E_{n,f}^2-\frac{q_0^2}{4})}\frac{(-2)}{e^{E_{n,f}/T}+1}.
\ee
Obviously, $I'_{2,vac}(m^2,q_0^2,0)$ is divergent and
in terms of the Pauli-Villars scheme, it takes the form ($q^0>0$)
\ba \label{eq:I2PV:vac:eB}
I'^{PV}_{2,vac}(m^2,q_0^2\pm i\epsilon,0)
&=&\sum_{\alpha=0}^N\frac{C_{\alpha}}{2\pi}\Bigg\{\Theta(2m_{n,f,\alpha}-q_0)
\Big[-\frac{\arcsin(\frac{q_0}{2m_{n,f,\alpha}})}{q_0m_{n,f,\alpha}\sqrt{1-(\frac{q_0}{2m_{n,f,\alpha}})^2}}\Big] \nonumber \\
&&+
\Theta(q_0-2m_{n,f,\alpha})\Big[\frac{\text{arccosh}(\frac{q_0}{2m_{n,f,\alpha}})\mp i\frac{\pi}{2}}{q_0m_{n,f,\alpha}\sqrt{(\frac{q_0}{2m_{n,f,\alpha}})^2-1}}\Big]\Bigg\},
\ea
where $\Theta(x)$ is the unit step function.
As for the temperature part $I'^{PV}_{2,tem}(m^2,q_0^2,0)$, it is easily obtained by
\be \label{eq:I2PV:tem:eB}
I'^{PV}_{2,tem}(m^2,q_0^2,0)=\sum_{\alpha=0}^N C_{\alpha}I'_{2,tem}(m_{\alpha}^2,q_0^2,0).
\ee
Thus, we have
\be \label{eq:I2PV:tot:eB}
I'^{PV}_{2}(m^2,q_0^2,0)=I'^{PV}_{2,vac}(m^2,q_0^2,0)+I'^{PV}_{2,tem}(m^2,q_0^2,0),
\ee
which can be used to calculate the pole mass of $\pi^0$ at finite $T$ and $eB$ by Eq.~(\ref{Gappole-re}).

Next, we show the equations for the screening masses of $\pi^0$ in the longitudinal direction, i.e. the direction of the magnetic field. In the frame of $q^{\mu}=(0,0,0,q^3)$, one finds that the polarization function for $\pi^0$ reads
\be\label{eq:PI:pi0:scr3:eB}
\Pi_{\pi^0}(0,q_3^2)=N_c\left[2I_1(m^2)+q_3^2I_2(m^2,0,q_3^2)\right].
\ee
Similarly, at $T=0$, $I_2(m^2,0,q_3^2)$ is represented in such a form
\be \label{eq:I2B:scr3:vac}
I_{2,vac}(m^2,0,q_3^2)=\sum_{f=u,d}\frac{|Q_f eB|}{2\pi} \sum_{n=0}^{\infty} \beta_n I'_{2,vac}(m^2,0,q_3^2),
\ee
with
\be \label{eq:I2PV:scr3:vac:eB}
I'_{2,vac}(m^2,0,q_3^2)
=i\int\frac{dp_0 dp_3}{(2\pi)^2} \frac{1}{p_0^2-p_3^2-m_{n,f}^2}\frac{1}{p_0^2-(p_3-q_3)^2-m_{n,f}^2}.
\ee
In the Pauli-Villars scheme, when making a replacement $q_3^2\rightarrow -(k_3^2\pm i\epsilon)$, $I'^{PV}_{2,vac}(m^2,0,k_3^2)$ takes the form
\ba \label{eq:I2PV:scr3:vac:eB}
I'^{PV}_{2,vac}(m^2,0, k_3^2\pm i\epsilon)
&=&\sum_{\alpha=0}^N\frac{C_{\alpha}}{2\pi}\Bigg\{\Theta(2m_{n,f,\alpha}-k_3)
\Big[-\frac{\arcsin(\frac{k_3}{2m_{n,f,\alpha}})}{k_3m_{n,f,\alpha}\sqrt{1-(\frac{k_3}{2m_{n,f,\alpha}})^2}}\Big]
\nonumber \\
&&+
\Theta(k_3-2m_{n,f,\alpha})\Big[\frac{\text{arccosh}(\frac{k_3}{2m_{n,f,\alpha}})
\mp i\frac{\pi}{2}}{k_3m_{n,f,\alpha}\sqrt{(\frac{k_3}{2m_{n,f,\alpha}})^2-1}}\Big]\Bigg\}.
\ea
Notice that it is exactly the same as Eq.~(\ref{eq:I2PV:vac:eB}), in consistent with our expectation that $m_{pole}=m_{scr,3}$ at $T=0$ but $eB\neq 0$.

For finite $T$, it is known that,
\be \label{eq:I2:scr3:tot:eB}
I'_{2}(m^2,0, q_3^2)=-T\sum^{\infty}_{l=-\infty} \int \frac{d p_3}{2\pi}\frac{1}{p_3^2+m_{n,l,f}^2}\frac{1}{(p_3-q_3)^2+m_{n,l,f}^2},
\ee
where $m_{n,l,f}^2=[(2l+1)\pi T]^2+m_{n,f}^2$. In order to evaluate the temperature-cut contribution easily, we use the method in Ref.~\cite{Ishii:2013kaa}, where the Matsubara summation must be taken after the momentum integration. Therefore, following the Pauli-Villars scheme, we have
\begin{eqnarray}\label{eq:I2PV:scr3:tot:eB}
I'^{PV}_{2}(m^2,0, q_3^2) &=&-\sum_{\alpha=0}^N C_{\alpha} \Bigg[T\sum^{\infty}_{l=-\infty} \int \frac{d p_3}{2\pi}\frac{1}{p_3^2+m_{n,l,f,\alpha}^2}\frac{1}{(p_3-q_3)^2+m_{n,l,f,\alpha}^2} \Bigg] \\ \nonumber
 &=&  -T\sum^{\infty}_{l=-\infty}\sum_{\alpha=0}^N \frac{C_{\alpha}}{4m_{n,l,f,\alpha}(m_{n,l,f,\alpha}^2+\frac{q_3^2}{4})},
\end{eqnarray}
with $m_{n,l,f,\alpha}^2=[(2l+1)\pi T]^2+m_{n,f,\alpha}^2$. Thus, after making the substitution $q_3^2\rightarrow -k_3^2$, we obtain
\be\label{eq:I2PV:scr3:tot123:eB}
I'^{PV}_{2}(m^2,0, k_3^2) = -T\sum^{\infty}_{l=-\infty}\sum_{\alpha=0}^N \frac{C_{\alpha}}{4m_{n,l,f,\alpha}(m_{n,l,f,\alpha}^2-\frac{k_3^2}{4})},
\ee
In this way, the longitudinal screening mass $m_{scr,\parallel}$ (i.e., $m_{scr,3}$) at finite $T$ and $eB$ is determined by
 \be\label{RPA-pion-long}
1-2G \, \text{Re}[\Pi^{PV}_{\pi^0}(0,\,k_3^2=m^2_{scr,\parallel})]=0
\ee

Finally, we turn to explore the screening masses of $\pi^0$ in the transverse direction $m_{scr,\perp}$ ($m_{scr,\perp}=m_{scr,1}=m_{scr,2}$) under the magnetic field along the positive z-axis. However, it is found that it is not convenient to compute the polarization functions by using the quark propagator in the LLR in momentum space of Eq.~(\ref{GDn-new}). Hence,
we make use of the quark propagator in the LLR in the mixed coordinate-momentum space of Eq.~(\ref{quark-mixed-GB2}), and the polarization function for $T=0$ can be expressed as (we set $\mathbf{r}_\perp'=0$ for simplicity)
 \be
 \Pi_{\pi^0,vac}(q)=-i\int\frac{dp_0 dp_3}{(2\pi)^2}\int d^2 \mathbf{r}_\perp \, e^{-i \mathbf{r}_\perp\cdot \mathbf{q}_\perp} \, \text{Tr}[i\gamma_5\tau^3\widetilde{S}(p_0,p_3;\mathbf{r}_\perp)i\gamma_5\tau^3
 \widetilde{S}(p_0-q_0,p_3-q_3;-\mathbf{r}_\perp)].
 \ee
In order to calculate $m_{scr,\perp}$, we need to set $q_0=q_3=0$. With the help of Eqs.~(\ref{A1}) and (\ref{A2}), we obtain
\begin{eqnarray}\label{PI-pion-perp-LL}
 \Pi_{\pi^0,vac}( 0, q_{\perp}^2)&=&N_c \sum_{f=u,d} \sum^{\infty}_{n,n'=0}\Bigg \{
 \frac{1}{2}(X_{n,n'}+X_{n-1,n'-1})[I'_{1,vac}(m_{n,f}^2)+I'_{1,vac}(m_{n',f}^{2})]\nonumber \\
 &&+\Big[\frac{(2n+2n')|Q_feB|}{2}
 \left(X_{n,n'}+X_{n-1,n'-1}\right)-Y_{n-1,n'-1}\Big]
 I''_{2,vac}(m^2)\Bigg \},
\end{eqnarray}
where $q_{\perp}^2= \mathbf{q}_\perp^2=q_1^2+q_2^2$, $X_{n,n'}=\frac{4\pi}{(2\pi l^2)^2}\mathcal{I}_{0}^{n,n'}(q_{\perp}^2)$
, $Y_{n,n'}=\frac{8\pi}{(2\pi l^2)^2l^4}\mathcal{I}_{2}^{n,n'}(q_{\perp}^2)$ (the explicit expressions of $\mathcal{I}_{0}^{n,n'}(q_{\perp}^2)$ and $\mathcal{I}_{2}^{n,n'}(q_{\perp}^2)$ are given in Appendix~\ref{appA}) and
\begin{eqnarray}
  I''_{2,vac}(m^2) &=& i\int\frac{dp_0dp_3}{(2\pi)^2}\frac{1}{(p_0^2-p_3^2-2n|Q_feB|-m^2)(p_0^2-p_3^2-2n'|Q_feB|-m^2)} \nonumber \\
  &=& \Bigg\{ \begin{array}{c}
             -\frac{1}{2\pi} \frac{\ln(m_{n',f}/m_{n,f})}{m_{n',f}^2-m_{n,f}^2}\,\,\,\,\, n'\neq n\\
             -\frac{1}{2\pi} \frac{1}{2m_{n,f}^2} \,\,\,\,\, n'= n.
           \end{array}
\end{eqnarray}
Thus, the corresponding expression in the Pauli-Villars scheme is given by
\be
 I''^{PV}_{2,vac}(m^2)=\sum_{\alpha=0}^N C_{\alpha} I''_{2,vac}(m_{\alpha}^2).
\ee
At finite $T$, $\Pi_{\pi^0}( 0, q_{\perp}^2)$ can be easily obtained by making the replacements $I'_{1,vac}(m^2)\rightarrow I'_{1}(m^2)$ and $I''_{2,vac}(m^2)\rightarrow I''_{2}(m^2)=I''_{2,vac}(m^2)+I''_{2,tem}(m^2)$, i.e.
\begin{eqnarray}\label{PI-pion-perp-LL-T}
 \Pi_{\pi^0}( 0, q_{\perp}^2)&=&N_c \sum_{f=u,d} \sum^{\infty}_{n,n'=0}\Bigg \{
 \frac{1}{2}(X_{n,n'}+X_{n-1,n'-1})[I'_{1}(m_{n,f}^2)+I'_{1}(m_{n',f}^{2})]\nonumber \\
 &&+\Big[\frac{(2n+2n')|Q_feB|}{2}
 \left(X_{n,n'}+X_{n-1,n'-1}\right)-Y_{n-1,n'-1}\Big]
 I''_{2}(m^2)\Bigg \},
\end{eqnarray}
where the temperature part $I''_{2,tem}(m^2)$ takes the form
\begin{eqnarray}
  I''_{2,tem}(m^2) &=& -I''_{2,vac}(m^2)+I''_{2}(m^2)\nonumber \\
  &=&-I''_{2,vac}(m^2)-T\sum^{\infty}_{l=-\infty}\int\frac{dp_3}{2\pi}
  \frac{1}{[(i\omega_l)^2-p_3^2-2n|Q_feB|-m^2][(i\omega_l)^2-p_3^2-2n'|Q_feB|-m^2]} \nonumber \\
  &=& \Bigg\{ \begin{array}{c}
             -\int\frac{dp_3}{2\pi} [\frac{1}{2E_{n,f}(E_{n,f}^2-E_{n',f}^2)}\frac{2}{e^{E_{n,f}/T}+1}
             +\frac{1}{2E_{n',f}(E_{n',f}^2-E_{n,f}^2)}\frac{2}{e^{E_{n',f}/T}+1}]\,\,\, \,\, n'\neq n\\
             -\int\frac{dp_3}{2\pi} \frac{1}{4E_{n,f}^3}\frac{-2}{e^{E_{n,f}/T}+1} \,\,\,\,\, n'= n.
           \end{array}
\end{eqnarray}
Similarly, we could have
\be
 I''^{PV}_{2,tem}(m^2)=\sum_{\alpha=0}^N C_{\alpha} I''_{2,tem}(m_{\alpha}^2).
\ee
And then by replacing $q_{\perp}^2\rightarrow -k_{\perp}^2$, the screening mass of $\pi^0$ in the transverse direction $m_{scr,\perp}$ can
be solved by
\be\label{RPA-pion-perp}
 1-2G\,\text{Re}[\Pi^{PV}_{\pi^0}(0,k_{\perp}^2=m_{scr,\perp}^2)]=0.
\ee

On the other hand, we find that it is more convenient to
obtain the transverse screening masses of $\pi^0$ by using the quark propagator in the PTR. More explicitly, for $T=0$, after substituting Eq.~(\ref{quark-pt-GB}) into Eq.~(\ref{pionpolar}) and performing straightforward but tedious calculations, one finds that $\Pi_{\pi^0,vac}(0,q_{\perp}^2)$ can be reduced to the following form
\ba\label{PI-meson-perp-PT-vac}
\Pi_{\pi^0,vac}(0,q_{\perp}^2)&=&\frac{N_{c}}{4 \pi^{2}} \sum_{f=u, d} \int_0^{\infty} d s \int_{0}^{1} d u \, e^{-s\left\{m^{2}+\frac{\sinh[Q_feB(\frac{1+u}{2})s]\sinh[Q_feB(\frac{1-u}{2})s]}{\sinh(Q_feBs)} \frac{q_{\perp}^2}{Q_feBs} \right\}}\\ \nonumber
&&\times\Bigg\{\left( m^{2}+\frac{1}{s}\right) \frac{Q_feB}{\tanh (Q_feBs)}+\frac{(Q_feB)^{2}}{\sinh ^{2} (Q_feBs)}-q_{\perp}^2\frac{Q_feB\sinh[Q_feB(\frac{1+u}{2})s]\sinh[Q_feB(\frac{1-u}{2})s]}{\sinh^3(Q_feBs)} \Bigg\}.
\ea
Of course, it is still divergent, but its convergent part contributed from the magnetic field could be easily extracted by subtracting the vacuum part. Thus, we rewrite $\Pi_{\pi^0,vac}(0,q_{\perp}^2)$ in the following form
\be\label{PI-meson-perp-PT2-vac}
\Pi_{\pi^0,vac}(0,q_{\perp}^2)=\Delta\Pi_{\pi^0,vac}(0,q_{\perp}^2)+\Pi_{\pi^0,vac}(0,q_{\perp}^2)|_{B= 0},
\ee
where
\ba\label{PI-meson-perp-PT3-vac}
\Delta\Pi_{\pi^0,vac}(0,q_{\perp}^2)
&=&\Pi_{\pi^0,vac}(0,q_{\perp}^2)-\Pi_{\pi^0,vac}(0,q_{\perp}^2)|_{B\rightarrow 0}\nonumber \\
&=&\frac{N_{c}}{4 \pi^{2}} \sum_{f=u, d} \int_0^{\infty} d s \int_{0}^{1} d u \, e^{-s\left\{m^{2}+\frac{\sinh[Q_feB(\frac{1+u}{2})s]\sinh[Q_feB(\frac{1-u}{2})s]}{\sinh(Q_feBs)} \frac{q_{\perp}^2}{Q_feBs} \right\}} \nonumber \\
&&\times\Bigg\{\left( m^{2}+\frac{1}{s}\right) \frac{Q_feB}{\tanh (Q_feBs)}+\frac{(Q_feB)^{2}}{\sinh ^{2} (Q_feBs)}-q_{\perp}^2\frac{Q_feB\sinh[Q_feB(\frac{1+u}{2})s]\sinh[Q_feB(\frac{1-u}{2})s]}{\sinh^3(Q_feBs)} \Bigg\} \nonumber \\
&&-\frac{N_{c}N_f}{4 \pi^{2}}\int_0^{\infty} d s \int_{0}^{1} d u \, e^{-s\left(m^{2}+\frac{1-u^2}{4} q_{\perp}^2 \right)}
\times\left( \frac{m^{2}}{s}+\frac{2}{s^2} -q_{\perp}^2\frac{1-u^2}{4s}  \right),
\ea
which is the finite contribution from the magnetic field. Note that the divergent part $\Pi_{\pi^0,vac}(0,q_{\perp}^2)|_{B= 0}$ need to be acquired by using the quark propagator in vacuum, and it has the form
\be \label{PI-meson-perp-vac-B0}
\Pi_{\pi^0,vac}(0,q_{\perp}^2)|_{B=0}=
N_c\left[2I_{1,vac}(m^2)|_{B=0}+q_{\perp}^2I_{2,vac}(m^2,0,q_{\perp}^2)|_{B=0}\right],
\ee
where $I_{2,vac}(m^2,0,q_{\perp}^2)|_{B=0}$ is defined by
\be \label{I2-meson-perp-vac-B0}
I_{2,vac}(m^2,0,q_{\perp}^2)|_{B=0}=2i N_f \int\frac{d^4p}{(2\pi)^4} \frac{1}{p_0^2-\mathbf{p}_{\perp}^2-p_3^2-m^2}\frac{1}{p_0^2-(\mathbf{p}_{\perp}-\mathbf{q}_{\perp})^2-p_3^2-m^2}.
\ee
And then, both terms of the right-hand side of Eq.~(\ref{PI-meson-perp-PT2-vac}) are handled within the Pauli-Villars regularization scheme,
\be\label{PI-meson-perp-PT2-PV-vac}
\Pi^{PV}_{\pi^0,vac}(0,q_{\perp}^2)=\Delta\Pi^{PV}_{\pi^0,vac}(0,q_{\perp}^2)
+\Pi^{PV}_{\pi^0,vac}(0,q_{\perp}^2)|_{B=0}.
\ee
The regularized form $\Delta\Pi^{PV}_{\pi^0,vac}(0,q_{\perp}^2)$ is obtained by making the substitution according to Eq.~(\ref{eq:PV}). As for $\Pi^{PV}_{\pi^0,vac}(0,q_{\perp}^2)|_{B=0}$, $I_{1,vac}^{PV}(m^2)|_{B=0}$ has been given by Eq.~(\ref{I1-PV-vac-B0}), and by following the method in Ref.~\cite{Florkowski:1997pi}, one finds the expression for $I_{2,vac}^{PV}(m^2,0,q_{\perp}^2)|_{B=0}$,
\ba \label{I2-meson-perp-vac-PV1-B0}
I_{2,vac}^{PV}(m^2,0,q_{\perp}^2)|_{B=0}
&=& \frac{N_f}{4\pi^2}\sum_{\alpha=0}^N C_{\alpha}\Bigg[\frac{2m_{\alpha}}{q_{\perp}}\sqrt{1+\Big(\frac{q_{\perp}}{2m_{\alpha}} \Big)^2}\ln\left(\sqrt{1+\Big(\frac{q_{\perp}}{2m_{\alpha}} \Big)^2}+ \frac{q_{\perp}}{2m_{\alpha}} \right)+\ln m_{\alpha}  \Bigg].
\ea
Moreover, the substitution $q_{\perp}^2\rightarrow -(k_{\perp}^2\pm i \epsilon)$ leads to the result ($k_{\perp}>0$)
\ba \label{I2-meson-perp-vac-PV2-B0}
I_{2,vac}^{PV}(m^2,0,k_{\perp}^2\pm i \epsilon))|_{B=0}
&=& \frac{N_f}{4\pi^2}\sum_{\alpha=0}^N C_{\alpha}\Bigg\{\Theta(2m_{\alpha}-k_{\perp})\Bigg[\frac{2m_{\alpha}}{k_{\perp}}\sqrt{1
-\Big(\frac{k_{\perp}}{2m_{\alpha}} \Big)^2}\arcsin\Big(\frac{k_{\perp}}{2m_{\alpha}} \Big)+\ln m_{\alpha}  \Bigg] \nonumber \\
&&+\Theta(k_{\perp}-2m_{\alpha})\Bigg[\frac{2m_{\alpha}}{k_{\perp}}\sqrt{\Big(\frac{k_{\perp}}{2m_{\alpha}} \Big)^2-1}\left(\operatorname{arcosh}\Big(\frac{k_{\perp}}{2m_{\alpha}} \Big)\mp i\frac{\pi}{2}\right)+\ln m_{\alpha}  \Bigg]\Bigg\}.
\ea

As for $T\neq 0$, by the sum over Matsubara frequencies we generalize the expression for the pion polarization function to the case of finite temperature, and the result for $\Pi_{\pi^0}(0,q_{\perp}^2)$ is given by
\ba\label{PI-meson-perp-PT-T}
\Pi_{\pi^0}(0,q_{\perp}^2)&=&\frac{N_{c}}{2 \pi^{\frac{3}{2}}}  \sum_{f=u, d} T \sum_{l=-\infty}^{\infty} \int_0^{\infty} s^{\frac{1}{2}} d s \int_{0}^{1} d u \, e^{-s\left\{m^{2}+\frac{\sinh[Q_feB(\frac{1+u}{2})s]\sinh[Q_feB(\frac{1-u}{2})s]}{\sinh(Q_feBs)} \frac{q_{\perp}^2}{Q_feBs}+\omega_l^2 \right\}} \nonumber\\
&&\times\Bigg\{\left(m^{2}+\frac{1}{2s}+\omega_l^2\right) \frac{Q_feB}{\tanh (Q_feBs)}
+\frac{(Q_feB)^{2}}{\sinh ^{2} (Q_feBs)} -q_{\perp}^2\frac{Q_feB\sinh[Q_feB(\frac{1+u}{2})s]\sinh[Q_feB(\frac{1-u}{2})s]}{\sinh^3(Q_feBs)} \Bigg\}.\nonumber\\
\ea
Similarly, it can be divided into two parts also,
\be\label{PI-meson-perp-PT2-T}
\Pi_{\pi^0}(0,q_{\perp}^2)=\Delta\Pi_{\pi^0}(0,q_{\perp}^2)
+\Pi_{\pi^0}(0,q_{\perp}^2)|_{B=0},
\ee
where
\ba\label{PI-meson-perp-PT3-T}
\Delta\Pi_{\pi^0}(0,q_{\perp}^2)
&=&\frac{N_{c}}{2 \pi^{\frac{3}{2}}}  \sum_{f=u, d} T \sum_{l=-\infty}^{\infty} \int_0^{\infty} s^{\frac{1}{2}} d s \int_{0}^{1} d u \, e^{-s\left\{m^{2}+\frac{\sinh[Q_feB(\frac{1+u}{2})s]\sinh[Q_feB(\frac{1-u}{2})s]}{\sinh(Q_feBs)} \frac{q_{\perp}^2}{Q_feBs}+\omega_l^2 \right\}} \nonumber\\
&&\times\Bigg\{\left(m^{2}+\frac{1}{2s}+\omega_l^2\right) \frac{Q_feB}{\tanh (Q_feBs)}
+\frac{(Q_feB)^{2}}{\sinh ^{2} (Q_feBs)} -q_{\perp}^2\frac{Q_feB\sinh[Q_feB(\frac{1+u}{2})s]\sinh[Q_feB(\frac{1-u}{2})s]}{\sinh^3(Q_feBs)} \Bigg\}\nonumber \\
&&-\frac{N_{c}N_f}{2 \pi^{\frac{3}{2}}}   T \sum_{l=-\infty}^{\infty} \int_0^{\infty} s^{-\frac{1}{2}} d s \int_{0}^{1} d u \, e^{-s\left\{m^{2}+ \frac{1-u^2}{4}q_{\perp}^2+\omega_l^2 \right\}}\\ \nonumber
&&\times\Bigg\{\left(m^{2}+\frac{3}{2s}+\omega_l^2\right)
-q_{\perp}^2\frac{1-u^2}{4} \Bigg\},
\ea
and $\Pi_{\pi^0}(0,q_{\perp}^2)|_{B=0}$ has the form
\be \label{PI-meson-perp-T-B0}
\Pi_{\pi^0}(0,q_{\perp}^2)|_{B=0}=
N_c\left[2I_{1}(m^2)|_{B=0}+q_{\perp}^2I_{2}(m^2,0,q_{\perp}^2)|_{B=0}\right],
\ee
with
\be \label{I2-meson-perp-T-B0}
I_{2}(m^2,0,q_{\perp}^2)|_{B=0}=-2 N_f T \sum_{l=-\infty}^{\infty}\int\frac{d^3p}{(2\pi)^3} \frac{1}{\omega_l^2+\mathbf{p}_{\perp}^2+p_3^2+m^2}\frac{1}{\omega_l^2
+(\mathbf{p}_{\perp}+\mathbf{q}_{\perp})^2+p_3^2+m^2}.
\ee
And both terms of $\Pi_{\pi^0}(0,q_{\perp}^2)$ need to be regularized in the same way as $\Pi_{\pi^0,vac}(0,q_{\perp}^2)$,
\be\label{PI-meson-perp-PT2-PV-T}
\Pi^{PV}_{\pi^0}(0,q_{\perp}^2)=\Delta\Pi^{PV}_{\pi^0}(0,q_{\perp}^2)
+\Pi^{PV}_{\pi^0}(0,q_{\perp}^2)|_{B=0}.
\ee
Especially, for $\Pi^{PV}_{\pi^0}(0,q_{\perp}^2)|_{B=0}$, $I_{1}^{PV}(m^2)|_{B=0}$ is defined by Eq.~(\ref{I1-PV-B0}), and $I_{2}^{PV}(m^2,0,q_{\perp}^2)|_{B=0}$ is given by the expression (see Ref.~\cite{Ishii:2013kaa} for detailed calculations)
\be \label{I2-meson-perp-T-PV-B0}
I_{2}^{PV}(m^2,0,q_{\perp}^2)|_{B=0}=-\frac{N_f T }{4\pi}\sum_{\alpha=0}^N\sum_{l=-\infty}^{\infty}\frac{C_{\alpha}}
{q_{\perp}}\arctan\left(\frac{q_{\perp}}{m_{\alpha}}\right).
\ee

In fact, the meson screening masses in the longitudinal direction could be calculated by the quark propagator in the proper-time representation also. By following the same procedure as above, we get the expressions of the corresponding polarization functions at $T=0$ and $T\neq 0$, respectively,
\ba\label{PI-meson-long-PT-vac}
\Pi_{\pi^0,vac}(0,q_{3}^2)&=&\frac{N_{c}}{4 \pi^{2}} \sum_{f=u, d} \int_0^{\infty} d s \int_{0}^{1} d u \, e^{-s\left[m^{2}+\frac{(1-u^2)}{4} q_3^2 \right]} \nonumber\\
&&\times\Bigg[\left(m^{2}+\frac{1}{s}\right) \frac{Q_feB}{\tanh (Q_feBs)}+\frac{(Q_feB)^{2}}{\sinh ^{2} (Q_feBs)}-q_{3}^2\frac{(1-u^2)}{4}  \frac{Q_feB}{\tanh (Q_feBs)} \Bigg]
\ea
and
\ba\label{PI-meson-long-PT-T}
\Pi_{\pi^0}(0,q_{3}^2)&=&\frac{N_{c}}{2 \pi^{\frac{3}{2}}}  \sum_{f=u, d} T \sum_{l=-\infty}^{\infty} \int_0^{\infty} s^{\frac{1}{2}} d s \int_{0}^{1} d u \, e^{-s\left[m^{2}+\frac{(1-u^2)}{4} q_3^2+\omega_l^2 \right]} \nonumber\\
&&\times\Bigg[\left( m^{2}+\frac{1}{2s}+\omega_l^2\right) \frac{Q_feB}{\tanh (Q_feBs)}
+\frac{(Q_feB)^{2}}{\sinh ^{2} (Q_feBs)} -q_{3}^2\frac{(1-u^2)}{4}  \frac{Q_feB}{\tanh (Q_feBs)} \Bigg].
\ea
The regularization procedure proceeds in the Pauli-Villars scheme as well,
\be\label{PI-meson-long-PT2-PV-vac}
\Pi^{PV}_{\pi^0,vac}(0,q_{3}^2)=\Delta\Pi^{PV}_{\pi^0,vac}(0,q_{3}^2)
+\Pi^{PV}_{\pi^0,vac}(0,q_{3}^2)|_{B=0}
\ee
and
\be\label{PI-meson-long-PT2-PV-T}
\Pi^{PV}_{\pi^0}(0,q_{3}^2)=\Delta\Pi^{PV}_{\pi^0}(0,q_{3}^2)
+\Pi^{PV}_{\pi^0}(0,q_{3}^2)|_{B=0}.
\ee
Note that $\Pi^{PV}_{\pi^0,vac}(0,q_{3}^2)|_{B=0}$ and $\Pi^{PV}_{\pi^0}(0,q_{3}^2)|_{B=0}$ take the same forms as $\Pi^{PV}_{\pi^0,vac}(0,q_{\perp}^2)|_{B=0}$ and $\Pi^{PV}_{\pi^0}(0,q_{\perp}^2)|_{B=0}$.
Finally, after making the replacement $q_3^2\rightarrow -k_3^2$ and $q_{\perp}^2\rightarrow -k_{\perp}^2$ in Eqs.~(\ref{PI-meson-long-PT2-PV-T}) and (\ref{PI-meson-perp-PT2-PV-T}), pion screening masses in the longitudinal and transverse direction can be solved by
Eqs.~(\ref{RPA-pion-long}) and~(\ref{RPA-pion-perp}), respectively.

\section{Numerical results}

In Sec. II, we have obtained the equations for the pole and screening masses of $\pi^0$ by means of the full RPA approach in a hot and magnetized medium. Before performing the numerical calculations, we remark that, in Refs.~\cite{Fayazbakhsh:2012vr,Fayazbakhsh:2013cha,Wang:2017vtn}, the derivative expansion method was employed to calculate meson masses in the NJL model. And this method is equivalent to the RPA in the LME (i.e. RRPA), e.g. $I_2(m^2,q^2)\approx I_2(m^2,0)$ in vaccum, as discussed in Ref.~\cite{Klevansky:1992qe}, where $I_2(m^2,q^2)$ is regarded as a smooth function dependent on $q^2$. Hence, in this section we will further compare the results generated by the FRPA with those generated by the RRPA so as to find out how these two approximations agree with each other. Apparently, for the pole masses and the longitudinal screening masses of mesons in the LME, we can just simply set $I_2(m^2,q_0^2,0)\approx I_2(m^2,q_0^2\rightarrow0,0)$ and $I_2(m^2,0,q_3^2)\approx I_2(m^2,0,q_3^2\rightarrow0)$ in Eqs.~(\ref{eq:I2B:vac}) and (\ref{eq:I2B:scr3:vac}), respectively. Please note that, as discussed in the introduction,  $I_2(m^2,q_0^2\rightarrow0,0)\neq I_2(m^2,0,q_3^2\rightarrow0)$ at finite temperature, which was neglected in Refs.~\cite{Fayazbakhsh:2012vr,Fayazbakhsh:2013cha,Wang:2017vtn}.

 As for the meson transverse screening masses, we need to expand the corresponding polarization functions, either in the LLR or in the PTR, to linear order in $q_{\perp}^2$. Explicitly, according to the expressions of Eqs.~(\ref{XX}) and (\ref{YY}) in the terms of the LLR, we have
\begin{eqnarray}\label{XX}
 \frac{1}{2}(X_{n,n'}+X_{n-1,n'-1})=(-1)^{n+n'}\frac{|Q_feB|}{2\pi}\left[\beta_n\delta_{n',n}
 -J_1^{n,n'}\frac{q_{\perp}^2}{2|Q_feB|}\right]+O(q_{\perp}^4)
\end{eqnarray}
and
\begin{eqnarray}\label{YY}
 Y_{n-1,n'-1}=(-1)^{n+n'}\frac{4|Q_feB|^2}{\pi}\left[n\delta_{n',n}
 -J_2^{n,n'}\frac{q_{\perp}^2}{2|Q_feB|}\right]+O(q_{\perp}^4),
\end{eqnarray}
where
\ba
J_1^{n,n'}&=&(2n-1)\delta_{n',n-1}+4n\delta_{n',n}+(2n+1)\delta_{n',n+1},\,\,\, \text{for} \, n\geq 1 \nonumber \\
J_1^{0,0}&=&J_1^{0,1}=1,\,\,\,\text{for} \, n=0
\ea
and
\ba
J_2^{n,n'}&=&n(n-1)\delta_{n',n-1}+2n^2\delta_{n',n}+n(n+1)\delta_{n',n+1},\,\,\, \text{for} \, n\geq 1.
\ea
So we can rewrite
$\Pi_{\pi^0,vac}( 0, q_{\perp}^2)$ and $\Pi_{\pi^0}( 0, q_{\perp}^2)$ in the LME,
\begin{eqnarray}\label{PI-pion-perp-LL-LML}
 \Pi^{LME}_{\pi^0,vac}( 0, q_{\perp}^2)&=&N_c \sum_{f=u,d} \sum^{\infty}_{n,n'=0} (-1)^{n+n'}\frac{|Q_feB|}{2\pi}\Bigg \{
 \beta_n \delta_{n',n} \Big[I'_{1,vac}(m_{n,f}^2)+I'_{1,vac}(m_{n',f}^{2})\Big]  \nonumber \\
 &&+q_{\perp}^2\Big[-(n+n')J_1^{n,n'}+4J_2^{n,n'}\Big]
 I''_{2,vac}(m^2)\Bigg \}+O(q_{\perp}^4)
\end{eqnarray}
and
\begin{eqnarray}\label{PI-pion-perp-LL-LML}
 \Pi^{LME}_{\pi^0}( 0, q_{\perp}^2)&=&N_c \sum_{f=u,d} \sum^{\infty}_{n,n'=0} (-1)^{n+n'}\frac{|Q_feB|}{2\pi}\Bigg \{
 \beta_n \delta_{n',n} \Big[I'_{1}(m_{n,f}^2)+I'_{1}(m_{n',f}^{2})\Big]  \nonumber \\
 &&+q_{\perp}^2\Big[-(n+n')J_1^{n,n'}+4J_2^{n,n'}\Big]
 I''_{2}(m^2)\Bigg \}+O(q_{\perp}^4).
\end{eqnarray}
On the other hand, in the PTR, we expand the Eqs.~(\ref{PI-meson-perp-PT-vac}) and (\ref{PI-meson-perp-PT-T}) to order $q_{\perp}^2$ straightforwardly,
\ba\label{PI-meson-perp-PT-vac-LML}
\Pi^{LME}_{\pi^0,vac}(0,q_{\perp}^2)&=&\frac{N_{c}}{4 \pi^{2}} \sum_{f=u, d} \int_0^{\infty} d s \int_{0}^{1} d u \, e^{-sm^{2} }\times\Bigg\{\Big( m^{2}+\frac{1}{s}\Big) \frac{Q_feB}{\tanh (Q_feBs)}+\frac{(Q_feB)^{2}}{\sinh ^{2} (Q_feBs)} \nonumber \\
&&-q_{\perp}^2\left[\Big(m^{2}+\frac{1}{s}\Big) \frac{Q_feB}{\tanh (Q_feBs)}+\frac{2(Q_feB)^{2}}{\sinh ^{2} (Q_feBs)}\right]\frac{\sinh[Q_feB(\frac{1+u}{2})s]\sinh[Q_feB(\frac{1-u}{2})s]}{Q_feB\sinh(Q_feBs)} \Bigg\}+O(q_{\perp}^4),\nonumber\\
\ea
and
\ba\label{PI-meson-perp-PT-T-LML}
\Pi^{LME}_{\pi^0}(0,q_{\perp}^2)&=&\frac{N_{c}}{2 \pi^{\frac{3}{2}}}  \sum_{f=u, d} T \sum_{l=-\infty}^{\infty} \int_0^{\infty} s^{\frac{1}{2}} d s \int_{0}^{1} d u \, e^{-s(m^{2}+\omega_l^2 )}\times\Bigg\{\Big(m^{2}+\frac{1}{2s}+\omega_l^2\Big) \frac{Q_feB}{\tanh (Q_feBs)}
+\frac{(Q_feB)^{2}}{\sinh ^{2} (Q_feBs)} \nonumber\\
&& -q_{\perp}^2\left[ \Big(m^{2}+\frac{1}{2s}+\omega_l^2\Big) \frac{Q_feB}{\tanh (Q_feBs)}
+\frac{2(Q_feB)^{2}}{\sinh ^{2} (Q_feBs)} \right]\frac{\sinh[Q_feB(\frac{1+u}{2})s]\sinh[Q_feB(\frac{1-u}{2})s]}{Q_feB\sinh(Q_feBs)} \Bigg\}+O(q_{\perp}^4).\nonumber\\
\ea
And then, after introducing the Pauli-Villars regularization, the transverse screening masses of $\pi^0$ in RRPA can be calculated by Eqs.~(\ref{RPA-pion-perp}). In fact, the polarization functions for the longitudinal screening masses of mesons in the LME could be also expressed in the PTR by expanding the Eqs.~(\ref{PI-meson-long-PT-vac}) and (\ref{PI-meson-long-PT-T}),
\ba\label{PI-meson-long-PT-vac-LML}
\Pi^{LME}_{\pi^0,vac}(0,q_3^2)&=&\frac{N_{c}}{4 \pi^{2}} \sum_{f=u, d} \int_0^{\infty} d s \int_{0}^{1} d u \, e^{-sm^{2} }\times\Bigg\{\Big( m^{2}+\frac{1}{s}\Big) \frac{Q_feB}{\tanh (Q_feBs)}+\frac{(Q_feB)^{2}}{\sinh ^{2} (Q_feBs)} \nonumber \\
&&-q_{3}^2\left[\Big(m^{2}+\frac{2}{s}\Big) \frac{Q_feB}{\tanh (Q_feBs)}+\frac{(Q_feB)^{2}}{\sinh ^{2} (Q_feBs)}\right]\frac{(1-u^2)}{4}s \Bigg\}+O(q_{3}^4),\nonumber\\
\ea
and
\ba\label{PI-meson-long-PT-T-LML}
\Pi^{LME}_{\pi^0}(0,q_{3}^2)&=&\frac{N_{c}}{2 \pi^{\frac{3}{2}}}  \sum_{f=u, d} T \sum_{l=-\infty}^{\infty} \int_0^{\infty} s^{\frac{1}{2}} d s \int_{0}^{1} d u \, e^{-s(m^{2}+\omega_l^2 )}\times\Bigg\{\Big(m^{2}+\frac{1}{2s}+\omega_l^2\Big) \frac{Q_feB}{\tanh (Q_feBs)}
+\frac{(Q_feB)^{2}}{\sinh ^{2} (Q_feBs)} \nonumber\\
&& -q_{3}^2\left[ \Big(m^{2}+\frac{3}{2s}+\omega_l^2\Big) \frac{Q_feB}{\tanh (Q_feBs)}
+\frac{(Q_feB)^{2}}{\sinh ^{2} (Q_feBs)} \right]\frac{(1-u^2)}{4}s \Bigg\}+O(q_{3}^4).\nonumber\\
\ea

\subsection{Results at fixed $eB$}

\begin{figure}
 \centerline{\includegraphics[scale=0.4]{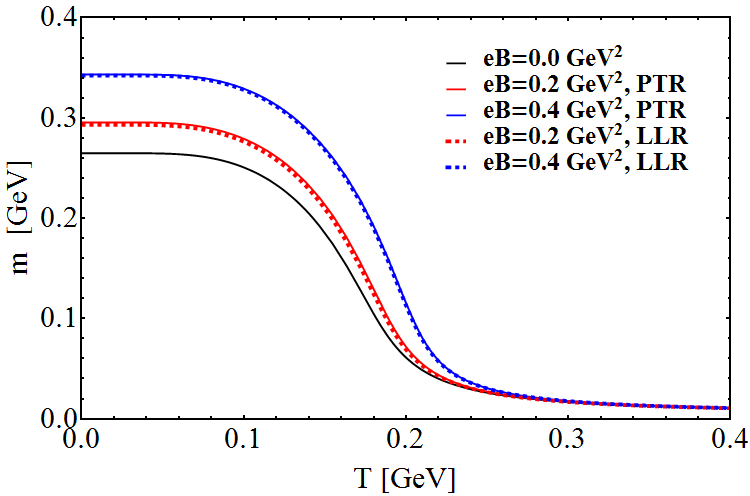}}
\caption{(Color online) Constituent quark mass $m$
as a function of T at $eB=0.0$, 0.2 and 0.4 $\text{GeV}^2$. For non-zero magnetic fields, i.e. $eB=0.2$ and 0.4 $\text{GeV}^2$, we show the comparison between the PTR and the LLR.}
\label{fig:gap:eB}
\end{figure}

By utilizing the gap equations in the form of either the LLR or the PTR, we first present the temperature dependence of the constituent mass $m$ for fixed $eB=0.0,\,0.2$ and $0.4$ $\text{GeV}^2$ in Fig.~\ref{fig:gap:eB}. For the LLR gap equation, we impose a sharp cutoff in the summation over the Landau level index at $n_{max}=1000$ for $eB=0.2$ and $0.4$ $\text{GeV}^2$, in order to achieve sufficient convergence. And it is shown that our numerical results obtained by the LLR formalism are almost the same as (only approximately 0.5 percent less than) the exact results obtained by the PTR formalism that contains the complete contribution of all Landau levels. Of course, it is easy to verify numerically that the more Landau levels are included in the LLR formalism, the smaller the discrepancy between these two formalisms is. Furthermore, from the Fig.~\ref{fig:gap:eB}, it can be seen that the values of $m$ (equivalent to the chiral condensate) increase with the magnetic field strength at any temperatures, which is the so called phenomenon of magnetic
catalysis. And as a consequence, the pseudo-critical temperature $T_c$ of the chiral phase transition increases with $eB$. More explicitly, we have $T_c(eB=0\,\text{GeV}^2)=173 \text{MeV}$, $T_c(eB=0.2\,\text{GeV}^2)=179 \text{MeV}$ and $T_c(eB=0.4\,\text{GeV}^2)=195 \text{MeV}$. Obviously, it is consistent with previous studies that the conventional NJL model gives rise to only  magnetic catalysis but no inverse magnetic catalysis. For simplicity, in this paper we will not take into account the effects of inverse magnetic catalysis and hope to address it in the future.

\begin{figure}
 \centerline{\includegraphics[scale=0.4]{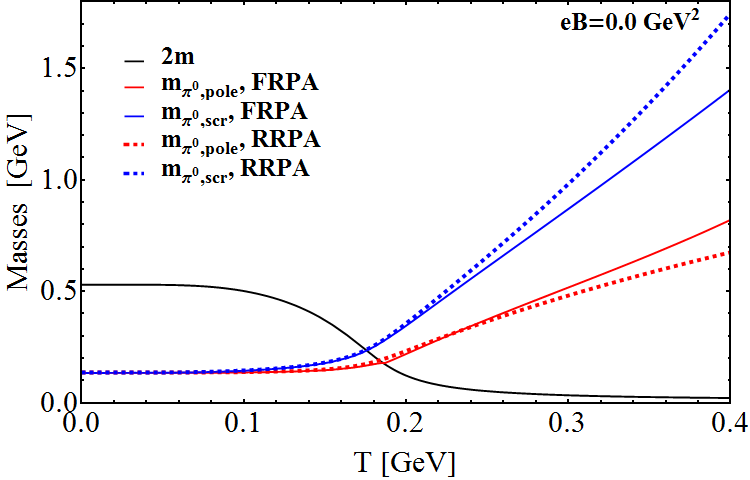}}
 \centerline{(a) }
 \smallskip
 \centerline{\includegraphics[scale=0.4]{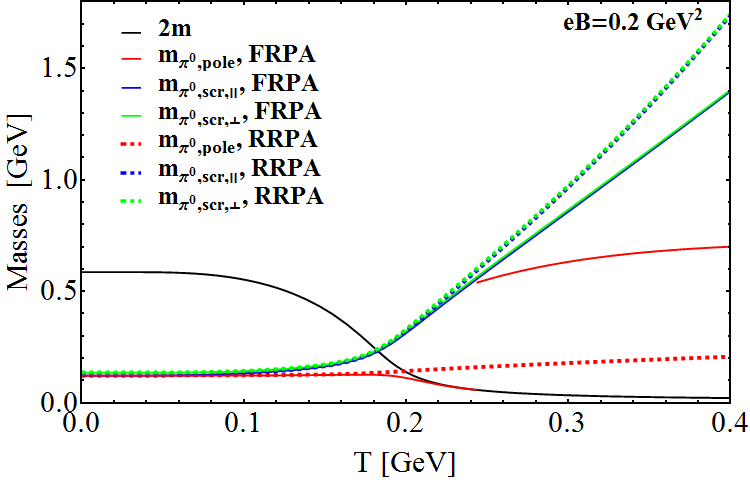}}
 \centerline{(b) }
 \smallskip
 \centerline{\includegraphics[scale=0.4]{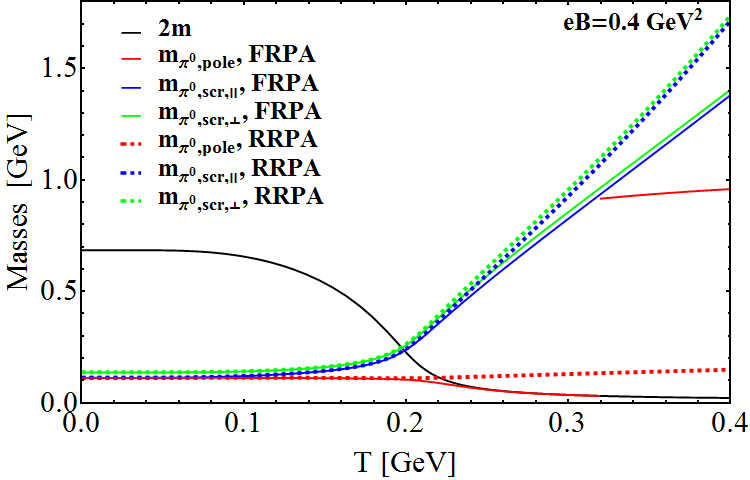}}
\centerline{(c)}
\caption{(color online) The $T$ dependence of $\pi^0$-meson pole masses $m_{\pi^0,pole}$, longitudinal screening masses $m_{\pi^0, scr, \parallel}$ and transverse screening masses $m_{\pi^0,scr,\perp}$, as well as $2m$, at $eB=0.0$, 0.2 and 0.4 $\text{GeV}^2$ within the FRPA and the RRPA. Especially, $m_{\pi^0, scr, \parallel}=m_{\pi^0,scr,\perp}=m_{\pi^0, scr}$ at $eB=0$.}
\label{fig:pion:eB}
\end{figure}

Now, we begin to show the pole mass, the longitudinal and transverse screening masses for $\pi^0$, as well as two times constituent quark mass $2m$, as functions of the temperature at fixed $eB=0.0$, $0.2$ and $0.4$ $\text{GeV}^2$ in Fig.~\ref{fig:pion:eB}. For comparison we present the results of pion masses generated by both the full RPA and the reduced RPA in this figure.
First of all, for $eB=0.0\,\text{GeV}^2$, as shown by the panel (a) of Fig.~\ref{fig:pion:eB}, the pion masses in the FRPA, including the pole and screening masses, remain small and approximately constant at low temperature ($T<T_c$). This is because of its nature as a pseudo-Goldstone boson in the
Nambu-Goldstone phase of chiral symmetry. When the temperature exceeds $T_c$, the chiral symmetry is partially restored. And it is found that both pole masses and screening masses of $\pi^0$ start to increase with the temperature remarkably in this Wigner-Weyl phase. Especially, when the pole mass $m_{\pi^0,pole}$ meets $2m$, it indicates the Mott transition temperature $T_{Mott}$ by the definition $m_{\pi^0,pole}(T_{Mott})=2m(T_{Mott})$, and $T_{Mott}(eB=0.0\,\text{GeV}^2)=186 \text{MeV}$ can be obtained. For the temperature larger than $T_{Mott}$, $\pi^0$ mesons become resonance states from bound states. On the other hand, $m_{\pi^0, scr}$ is always greater than $m_{\pi^0, pole}$ at non-zero temperatures due to the breaking of the Lorentz covariance by the heat bath. As the temperature increases, the mass splitting between $m_{\pi^0, pole}$ and $m_{\pi^0, scr}$ gets larger and larger, which means the enhancement of the symmetry breaking. And the remnant $SO(3)$ symmetry suggests $m_{\pi^0, scr, \parallel}=m_{\pi^0,scr,\perp}$ at finite $T$ and vanishing $eB$. Additionally, in the interval $0 < T < 250$ MeV, the results of $m_{\pi^0,pole}$, $m_{\pi^0, scr, \parallel}$ and $m_{\pi^0,scr,\perp}$ in the RRPA is almost the same as those in the FRPA. However, when $T > 250$ MeV, the meson mass is too heavy to make the LME method sufficient, since a smooth dependence on external momenta for the function $I_2$ is not valid any more.
And it shows in the figure that, for $T > 250$ MeV, $m_{\pi^0, pole}$ in the RRPA is smaller than those in the FRPA, while $m_{\pi^0, scr}$ in the RRPA is larger than those in the FRPA.

Next, we turn to the results at finite $eB$, i.e. $eB=0.2$ and $0.4$ $\text{GeV}^2$ in the panel (b) and (c) of Fig.~\ref{fig:pion:eB}, respectively. When at low temperatures, the behaviors of all three kinds of $\pi^0$ masses curves at non-vanishing magnetic fields are similar to those at $eB=0.0$ $\text{GeV}^2$. This is because the magnetic field helps to enhance the breaking of the chiral symmetry so that the $\pi^0$ mesons remain in the
Nambu-Goldstone phase at low temperature. In this temperature region, both pole masses and screening masses display almost the same behaviors either in the FRPA or in the RRPA.

However, in the Wigner-Weyl phase, the pole masses of $\pi^0$ in the FPRA  at $eB\neq 0$ show some difference from those at
$eB=0$. An important difference is the mass jump of the $m_{\pi^0, pole}$ at $T_{Mott}$, where the pole mass of $\pi^0$ suddenly jumps from nearly $2m$ to a more energetic state. The explanation for the mass jump of $m_{\pi^0, pole}$ has been thoughtfully discussed in Refs.~\cite{Mao:2017wmq} and \cite{Avancini:2018svs}. It is argued that the dimensional reduction associated with the magnetic fields
leads to an infrared divergence for the
lowest Landau level at the threshold $m_{\pi^0, pole}=2m$. As a consequence, the threshold mass $2m$ is not sufficiently to become the solution of the RPA equation of the $\pi^0$ pole mass, and it has to jumps to a more energetic state at the Mott transition temperature. On the other hand, the infrared divergence will enforce the $\pi^0$ pole mass to approach $2m$ infinitely when $T<T_{Mott}$. Such behaviors of $m_{\pi^0, pole}$ are shown in the panel (b) and (c) of Fig.~\ref{fig:pion:eB} also: between the interval of $T_c < T < T_{Mott}$ at $eB=0.2$ and $0.4$ $\text{GeV}^2$, $m_{\pi^0, pole}$ acquired by the FRPA increases slightly at first, and then decreases with temperature to get close to $2m$. Our results are consistent with the results found in Ref.~\cite{Avancini:2018svs}, but opposite to the results in Ref.~\cite{Mao:2017wmq}, where $m_{\pi^0, pole}$
grows with $T$ monotonically when $T < T_{Mott}$. This difference results in distinct effects of magnetic fields on the Mott temperature:
$T_{Mott}$ is catalyzed by the magnetic field in our paper and Ref.~\cite{Avancini:2018svs}, while it is anti-catalyzed by the magnetic field in Ref.~\cite{Mao:2017wmq}. Actually, in the scenario of the chiral limit, we have $T_{Mott}=T_c$, and it should increase with the magnetic field in the conventional NJL model. When considering the discrepancy between the FRPA and the RRPA for $m_{\pi^0, pole}$ at $eB \neq 0$, it is clear from the graphs in Fig.~\ref{fig:pion:eB} that $m_{\pi^0, pole}$ obtained within the RRPA do not show any mass jump and increase monotonously with the increasing of the temperature, since no infrared divergence appears in the equation of $m_{\pi^0, pole}$ within the LME. And more specifically,
in the interval of $T_c<T<T_{Mott}$, the results within the RRPA are higher than those within the FRPA, but while $T>T_{Mott}$, the results within the RRPA are much lower than those within the FRPA.

As for the temperature dependence of the
screening masses $m_{\pi^0, scr, \parallel}$ and $m_{\pi^0,scr,\perp}$ at $eB \neq 0$, it is found that they show the behaviors similar to those at $eB =0$, and do not show any mass jumps unlike $m_{\pi^0, pole}$. The main reason of this is that the vacuum and temperature-cut contributions partially cancel each other for the polarization function of the
screening masses at finite temperature, as discussed in Refs.~\cite{Ishii:2013kaa}, which make the expressions suffer no infrared divergence at any Landau levels, e.g.~Eq.~(\ref{eq:I2PV:scr3:tot123:eB}). Moreover, by comparing the results in the FRPA with the results in the RRPA for $m_{\pi^0, scr, \parallel}$ and $m_{\pi^0,scr,\perp}$ at $eB \neq 0$, the difference between these two prescriptions is qualitatively consistent with the situation of $eB=0$ as discussed above.

\begin{figure}
 \centerline{\includegraphics[scale=0.5]{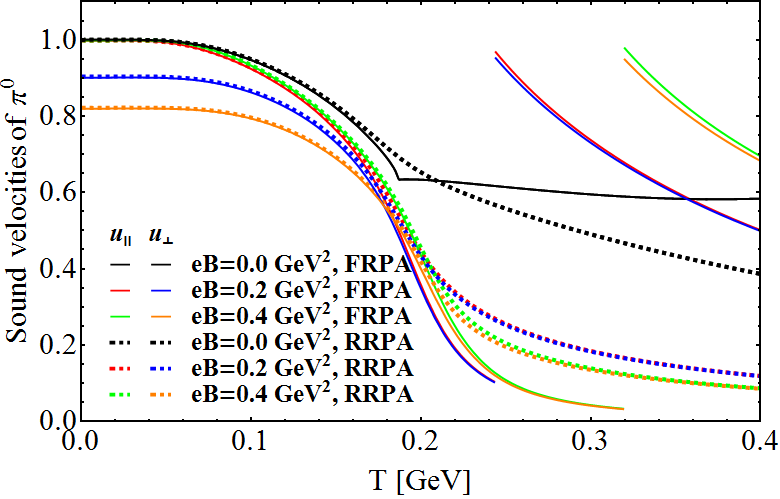}}
 \centerline{(a) }
 \smallskip
 \centerline{\includegraphics[scale=0.5]{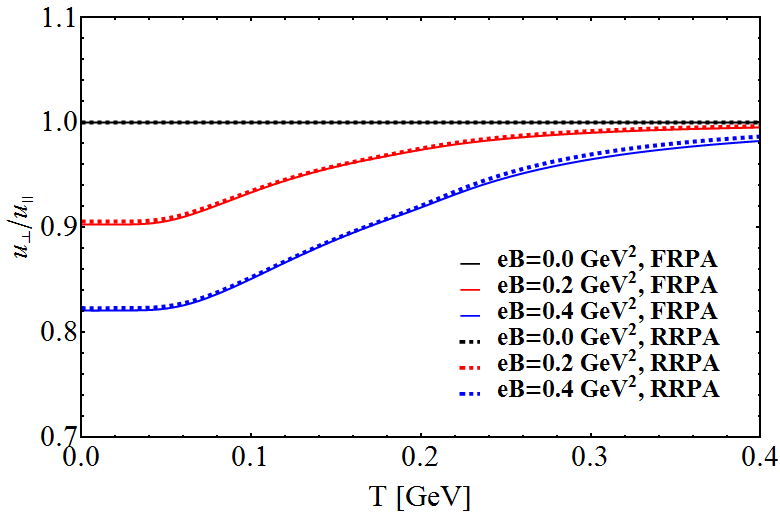}}
\centerline{(b)}
\caption{(color online) The $T$ dependence of longitudinal sound velocity $u_{\parallel}$ and transverse sound velocity $u_{\perp}$, as well as the ratio ${u_{\perp}}/{u_{\parallel}}$, at $eB=0.0$, 0.2 and 0.4 $\text{GeV}^2$ for $\pi^0$ within the FRPA and the RRPA. Especially, $u_{ \parallel}=u_{\perp}$ at $eB=0$.}
\label{fig:u:eB}
\end{figure}

And then, we sketch in Fig.~\ref{fig:u:eB} the temperature dependence of the sound velocities of $\pi^0$ including $u_{\parallel}$ and $u_{\perp}$, as well as the ratio ${u_{\perp}}/{u_{\parallel}}$, at fixed $eB=0.0$, $0.2$ and $0.4$ $\text{GeV}^2$, which are evaluated by the FRPA and the RRPA also.
Obviously, we find that, due to the mass jump of $m_{\pi^0, pole}$ at $eB \neq 0$, the behaviors of $u_{\parallel}$ and $u_{\perp}$ at $eB=0$ are quite different from those at $eB \neq 0$ in the FRPA. For $eB=0$, $u_{\parallel}$ (i.e. $u_{\perp}$) obtained by the FRPA, declines with temperature continuously but shows non-differentiability at $T_{Mott}$. In the high temperature limit ($T\rightarrow 400$ MeV), the sound velocities of $\pi^0$ approach $0.58\sim \frac{\sqrt{3}}{3}$, which corresponds to a gas of non-interacting quarks.
On the other hand, for results of $eB=0.2$ and $0.4$ $\text{GeV}^2$ in the FRPA, it is shown that, as $T$ increases, $u_{\parallel}$ decreases from unity first but jumps to about unity again at $T_{Mott}$, and then continues to show a decreasement; $u_{\perp}$ behaves in a similar way to $u_{\parallel}$, except that the starting points of it at $T=0$ decrease with $eB$, owing to the enhancement of the symmetry breaking ($SO(3)\rightarrow SO(2)$) in spatial space caused by the external magnetic field. As the results of $u_{\parallel}$ and $u_{\perp}$ in the RRPA, they all reduce with the increase of temperature continuously, since there is no mass jump for $m_{\pi^0, pole}$.

In fact, the results of the ratio ${u_{\perp}}/{u_{\parallel}}$ in the panel (b) of Fig.~\ref{fig:u:eB}, are not more than unity, as it is related to the relative refractive index $n_{\perp}(B,T)/n_{\parallel}(T)=u_{\parallel}/u_{\perp}$ of the medium and reflects the screening effect of the magnetic fields.
Besides, we can find the fact that, for $eB=0.2$ and $0.4$ $\text{GeV}^2$,
${u_{\perp}}/{u_{\parallel}}$ is temperature independent in the interval of $0<T<50$ MeV, and then increases with $T$ and approaches unity gradually when $T>50$ MeV. It is shown that only when $T<50$ MeV, the screening effect of the temperature can be decoupled from that of the magnetic field, and ${u_{\perp}}/{u_{\parallel}}$ solely depends on the magnetic field strength. But when $T>50$ MeV, the temperature will dilute the anisotropy
stemming from the magnetic field. In addition, although the deviation between the FRPA and the RRPA for $m_{\pi^0, scr, \parallel}$ and $m_{\pi^0,scr,\perp}$ becomes larger and larger when $T>250$ MeV, the ratios of two kinds of screening masses, i.e. $m_{\pi^0, scr, \parallel}/m_{\pi^0,scr,\perp}={u_{\perp}}/{u_{\parallel}}$, evaluated by the FRPA and the RRPA show agreement with each other in the whole temperature region.

In Refs.~\cite{Fayazbakhsh:2012vr,Fayazbakhsh:2013cha,Wang:2017vtn} the authors use the derivative expansion method in the NJL model to compute $m_{\pi^0, pole}$, $m_{\pi^0, scr, \parallel}$ and $m_{\pi^0,scr,\perp}$, as well as $u_{\parallel}$ and $u_{\perp}$, at finite $T$ for different fixed $eB$ within the RRPA.
On the one hand, it is shown that the values of $u_{\perp}$ at $eB\neq 0$ in Refs.~\cite{Fayazbakhsh:2012vr,Fayazbakhsh:2013cha} are always larger than the
speed of light, which thus violates the
law of causality. This is because that they made use of the noncovariant regularization scheme and it can be cured in the
covariant regularization schemes (e.g., the Pauli-Villars
regularization scheme), as discussed in Ref.~\cite{Mao:2017wmq}. On the other hand, the temperature dependence of $u_{\parallel}$ at $eB\neq 0$ is identically equal to unity in Ref.~\cite{Fayazbakhsh:2012vr,Fayazbakhsh:2013cha,Wang:2017vtn}, while our results of $u_{\parallel}$ in the RRPA decrease with the temperature, which implies breaking of the Lorentz invariance at finite $T$. Actually, this disagreement is caused by the defect of derivative expansion method, which neglects the difference between the static limit and the long-wavelength limit at finite temperature. Relying on our appropriate approaches, we solve the above problems and reasonable numerical results are displayed in our article.

\subsection{Results at fixed $T$}

\begin{figure}
 \centerline{\includegraphics[scale=0.4]{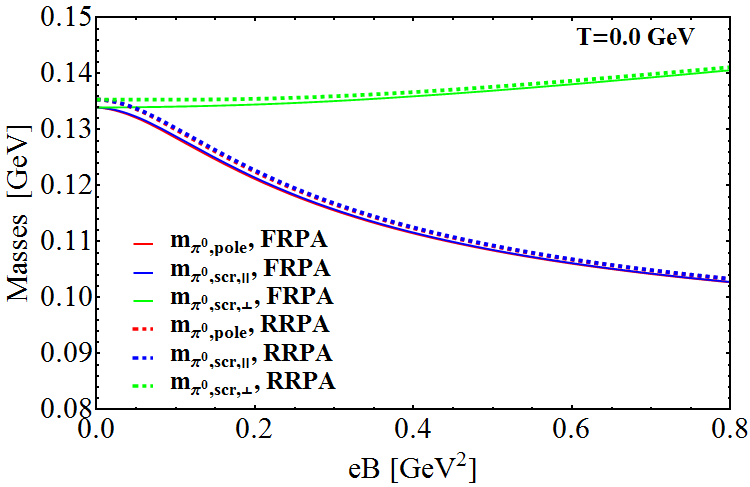}}
 \centerline{(a) }
 \smallskip
 \centerline{\includegraphics[scale=0.4]{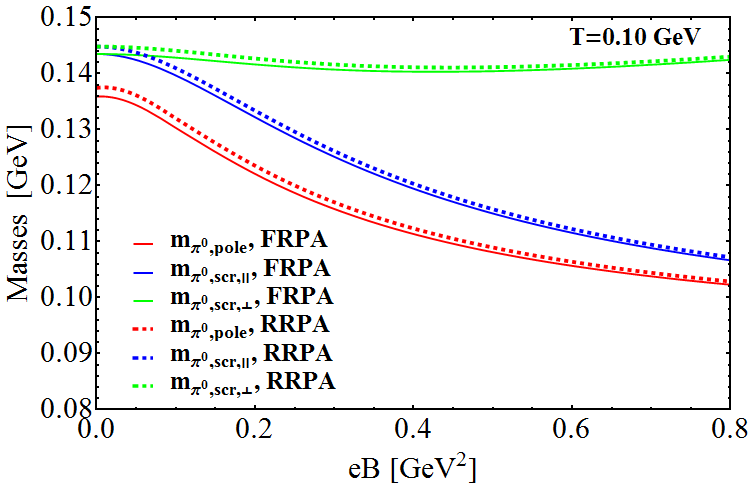}}
 \centerline{(b) }
 \smallskip
 \centerline{\includegraphics[scale=0.4]{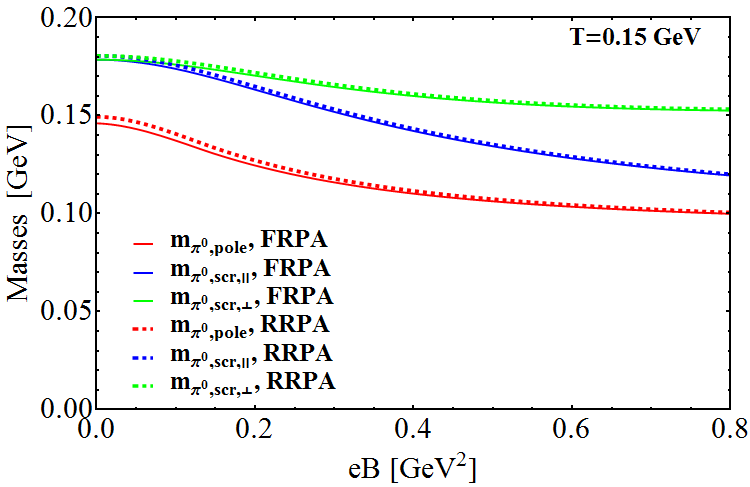}}
\centerline{(c)}
\caption{(color online) The $eB$ dependence of $\pi^0$-meson pole masses $m_{\pi^0,pole}$, longitudinal screening masses $m_{\pi^0, scr, \parallel}$ and transverse screening masses $m_{pi^0,scr,\perp}$ at $T=0.0$, 0.10 and 0.15 $\text{GeV}$ within the FRPA and the RRPA.}
\label{fig:pion:T}
\end{figure}

In this subsection, we present the $eB$ dependence of $m_{\pi^0, pole}$, $m_{\pi^0, scr, \parallel}$ and $m_{\pi^0,scr,\perp}$ at fixed temperature $T=0$, $0.10$ and $0.15$ GeV in Fig.~\ref{fig:pion:T}.
In order to avoid the ambiguity resulted from the mass jump of $m_{\pi^0, pole}$ at $T_{Mott}$, we focus on the temperature region below $T_c$, where
it is shown that the difference between the FRPA and the RPA extremely small. Firstly,
for $T=0$, as shown by the panel (a) of Fig.~\ref{fig:pion:T}, $m_{\pi^0, pole}$ decreases as the external magnetic field grows, which is in
agreement with the lattice results in Ref.~\cite{Ding:2020hxw}. And $m_{\pi^0, scr, \parallel}$ is identical to $m_{\pi^0, pole}$ at any magnetic field strength because of the Lorentz invariance between
the time direction and the magnetic field direction at zero
temperature, even though they are computed by the PTR and the LLR, respectively.
As concerns $m_{\pi^0,scr,\perp}$, it increase with $eB$ at $T=0$. And more importantly, the mass splitting between $m_{\pi^0,scr,\perp}$ and $m_{\pi^0, scr, \parallel}$ goes up with the increasing $eB$, which means that the breaking of the Lorentz invariance is enhanced by the increase of the magnetic fields.

Secondly, for $T=0.10$ and $0.15$ GeV, depicted by the panels (b) and (c) of Fig.~\ref{fig:pion:T}, $m_{\pi^0, pole}$ and $m_{\pi^0, scr, \parallel}$ still show the decreasing behaviors as $eB$ increases, but the mass splitting between them rises with the growth of $T$, which implies the breaking of the Lorentz invariance between the temporal
direction and the magnetic field direction. As for the $eB$ dependence of $m_{\pi^0,scr,\perp}$ at $T\neq 0$, the situation becomes different: for $T=0.10$ GeV, the curve of $m_{\pi^0,scr,\perp}$ nearly remains constant, as the magnetic field increases; but for $T=0.15$ GeV, $m_{\pi^0,scr,\perp}$ turns to decrease with the increasing $eB$. It shows that the decreasing behavior of $m_{\pi^0,scr,\perp}$ appears when the temperature is beyond a certain threshold temperature $T_0\approx 100$ MeV.

\begin{figure}
 \centerline{\includegraphics[scale=0.5]{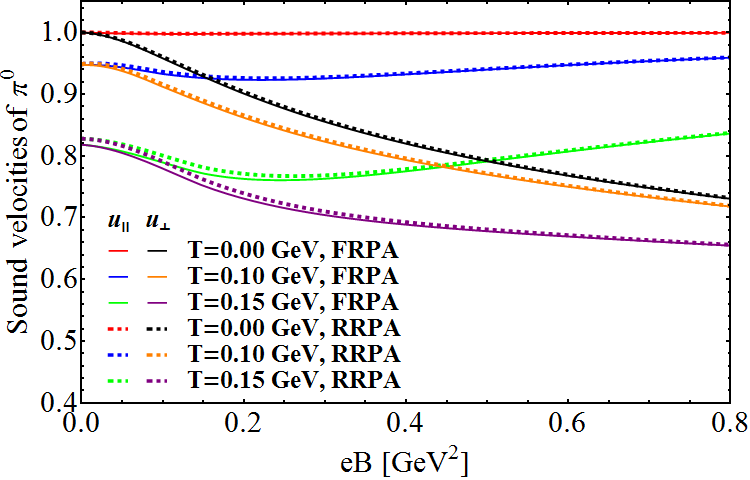}}
 \centerline{(a) }
 \smallskip
 \centerline{\includegraphics[scale=0.5]{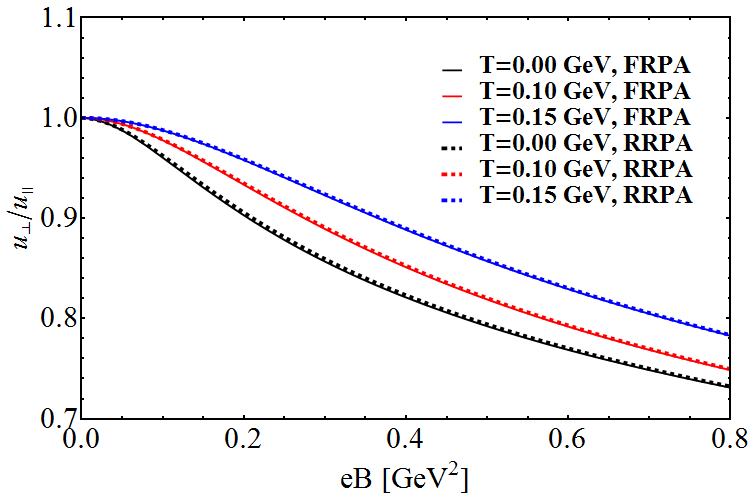}}
\centerline{(b)}
\caption{(color online) The $eB$ dependence of longitudinal sound velocity $u_{\parallel}$ and transverse sound velocity $u_{\perp}$, as well as the ratio ${u_{\perp}}/{u_{\parallel}}$, at $T=0.0$, 0.10 and 0.15 $\text{GeV}$ for $\pi^0$ within the FRPA and the RRPA.}
\label{fig:u:T}
\end{figure}

Furthermore, in Fig.~\ref{fig:u:T}, we plot the $u_{\parallel}$, $u_{\perp}$ and ${u_{\perp}}/{u_{\parallel}}$ as functions of $eB$ at $T=0$, $0.10$ and $0.15$ GeV. According to the Lorentz invariance between the temporal direction and the magnetic field direction, it is obvious that $u_{\parallel}$ always equals to the speed of light at $T=0$. And with respect to $T \neq 0$, when the temperature is low ($T\lesssim 0.1$ GeV), $u_{\parallel}$ shows smooth dependence on $eB$, but when the temperature is high enough, $u_{\parallel}$ first decreases and then
increases with the magnetic field strength $eB$ and seems to saturate at $eB>0.6$ $\text{GeV}^2$. More
explicitly, it means that the magnetic field will enhances first and then reduces the anisotropy between the temporal direction and the longitudinal direction caused by the temperature. Therefore, only in the low temperature region or strong magnetic field region, the behaviors of $u_{\parallel}$ are consistent with the expectation that $u_{\parallel}=u_{\parallel}(T)$. Otherwise, $u_{\parallel}$ shows dependence not only on the temperature but also on the magnetic field strength. As for $u_{\perp}$ and ${u_{\perp}}/{u_{\parallel}}$, they both continuously decline with $eB$, reflecting the enhancement of the anisotropy in coordinate space by the magnetic field. The main difference between them is that the starting points of $u_{\perp}$ at $eB=0$ decrease with $T$, while the starting points of ${u_{\perp}}/{u_{\parallel}}$ is equal to unity always as the temperature increases, which is related to the breaking of the Lorentz invariance by the heat bath. What's more, we can find that, the higher the temperature is, the slower the ratio ${u_{\perp}}/{u_{\parallel}}$ decreases with $eB$. It means that the increasing of the temperature will help to weaken the breaking of the Lorentz invariance by the magnetic field, as we have mentioned above.

\section{summary and conclusions}

In this work, we have presented a comprehensive and systematical study on the mass spectrum, including pole masses and directional screening masses, of neutral pions at finite temperature and magnetic field by the RPA approach in the NJL model, where two mathematically equivalent formalisms have been used (except the pole masses in the PTR): the Landau level representation (LLR) and the proper-time representation (PTR). In particular, for comparison we provide the numerical results both in the FRPA and in the RRPA. Additionally, the behaviors of the directional sound velocities $u_{\parallel}$ and $u_{\perp}$, as well as the ratio ${u_{\perp}}/{u_{\parallel}}$, are all investigated in the hot and magnetized medium.

By analyzing the $T$ dependence of $\pi^0$ masses at fixed $eB$, we find that in the low temperature regime ($T<T_c$), as the pseudo-Goldstone boson for chiral symmetry breaking, the masses of $\pi^0$ (either pole masses or directional screening masses) nearly maintain a constant value at any fixed magnetic field. However, the pole masses of $\pi^0$ at nonzero magnetic field experience a sudden mass jump at $T_{Mott}$ resulted from the dimensional reduction associated with the magnetic field. What's more, the
Mott temperature $T_{Mott}$ is catalyzed with the increase of the magnetic field as well as the pseudo-critical temperature $T_c$, which qualitatively coincides with the results in Ref.~\cite{Avancini:2018svs}, but contradicts with the results in Ref.~\cite{Mao:2017wmq}. And the mass jump of $m_{\pi^0, pole}$ in the FPRA at finite $eB$ also results in the jumps of $u_{\parallel}$ and $u_{\perp}$ at
$T_{Mott}$. By making use of the method in Ref.~\cite{Ishii:2013kaa}, the $T$ dependence of $m_{\pi^0, scr, \parallel}$ and $m_{\pi^0,scr,\perp}$ at finite magnetic field is calculated in our paper and exhibits no mass jump. On the other hand, the analysis of the $eB$
dependence of $\pi^0$ masses at fixed $T$ reveals that the magnetic field
strengthens the breaking of the Lorentz invariance between the longitudinal direction and transverse direction, while the temperature helps to recover the asymmetry caused by the magnetic field.

It should be emphasized that, in this study, we clarify the reason in detail why the directional sound velocities $u_{\parallel}$ and $u_{\perp}$, as well as the ratio ${u_{\perp}}/{u_{\parallel}}$ violate the law of causality in Refs.~\cite{Fayazbakhsh:2012vr,Fayazbakhsh:2013cha}, which was argued in Ref.~\cite{Mao:2017wmq} also. And by using the covariant
Pauli-Villars regularization scheme, reasonable behaviors are well described within either the FRPA or the RRPA in our paper, which show the screening effects of the temperature and the magnetic field. The heat bath leads to the asymmetry between the temporal direction and the spatial direction, while the magnetic field leads to the asymmetry between the longitudinal direction and transverse direction. Hence, we must have $m_{\pi^0, pole}<m_{\pi^0, scr, \parallel}<m_{\pi^0,scr,\perp}$, i.e.$u_{\parallel}<1$, $u_{\perp}<1$ and ${u_{\perp}}/{u_{\parallel}}<1$ at finite $T$ and $eB$.
And another point we need to notice is that, because of the limitation of the derivative expansion method at finite temperature, the values of $u_{\parallel}$ within the RRPA in Refs.~\cite{Fayazbakhsh:2012vr,Fayazbakhsh:2013cha,Wang:2017vtn} do not show the anisotropy resulted from the heat bath. We can only rely on taking the limits in the correct order within the FRPA to achieve the corresponding results.

\acknowledgements
The authors thank Igor A. Shovkovy, Mei Huang and Danning Li for useful discussion. L.Y. acknowledges the kind hospitality of the College of Integrative Sciences and Arts at Arizona State University, and the School of Nuclear Science and Technology at University of Chinese Academy of Sciences.
The work of L.Y. is supported by the
NSFC under Grant No. 11605072 and the Seeds Funding of Jilin University.
X.W. is supported by
the start-up funding No. 4111190010 of Jiangsu University.

\appendix

\section{Useful formulas}\label{appA}

Here we list the expressions of $\mathcal{I}_{0}^{n,n'}(\mathbf{k}_\perp^2)$ and $\mathcal{I}_{2}^{n,n'}(\mathbf{k}_\perp^2)$ used in Sec.II (the details of the derivations for these functions are given in Ref.~\cite{Pyatkovskiy:2010xz}):
\begin{eqnarray} \label{A1}
\mathcal{I}_{0}^{n,n'}(\mathbf{k}_\perp^2)&=&\int_{0}^{\infty} r_\perp d r_\perp e^{-\mathbf{r}_\perp^2/(2l^2)} J_{0}\left(r_{\perp}k_\perp\right)L_{n}\left(\frac{\mathbf{r}_{\perp}^2}{2l^{2}}\right)
L_{n^\prime}\left(\frac{\mathbf{r}_{\perp}^2}{2l^{2}}\right)\\ \nonumber
&=& (-1)^{n+n^\prime} l^2  e^{-\mathbf{k}_{\perp}^2l^{2}/2} L_{n}^{n^\prime-n}\left(\frac{\mathbf{k}_{\perp}^2l^{2}}{2}\right)
L_{n^\prime}^{n-n^\prime}\left(\frac{\mathbf{k}_{\perp}^2l^{2}}{2}\right) \\ \nonumber
&=& l^2 \frac{n_{<}!}{n_>!} e^{-\mathbf{k}_{\perp}^2l^{2}/2} \left(\frac{\mathbf{k}_{\perp}^2l^{2}}{2}\right)^{|n-n'|} \bigg[L_{n_<}^{|n-n'|}\left(\frac{\mathbf{k}_{\perp}^2l^{2}}{2}\right)\bigg]^2
,\\
\label{A2}
\mathcal{I}_{2}^{n,n'}(\mathbf{k}_\perp^2)&=&\int_{0}^{\infty} r_\perp^3  d r_\perp e^{-\mathbf{r}_\perp^2/(2l^2)} J_{0}\left(r_{\perp}k_\perp\right) L_{n}^{1}\left(\frac{\mathbf{r}_{\perp}^2}{2l^{2}}\right)
L_{n^\prime}^{1}\left(\frac{\mathbf{r}_{\perp}^2}{2l^{2}}\right)\\ \nonumber &=&
 2 (-1)^{n+n^\prime} l^4 (n^\prime+1)
e^{-\mathbf{k}_{\perp}^2l^{2}/2} L_{n}^{n^\prime-n}\left(\frac{\mathbf{k}_{\perp}^2l^{2}}{2}\right)
L_{n^\prime+1}^{n-n^\prime}\left(\frac{\mathbf{k}_{\perp}^2l^{2}}{2}\right) \\ \nonumber
&=& 2l^4 \frac{(n_{<}+1)!}{n_>!} e^{-\mathbf{k}_{\perp}^2l^{2}/2} \left(\frac{\mathbf{k}_{\perp}^2l^{2}}{2}\right)^{|n-n'|} L_{n_<}^{|n-n'|}\left(\frac{\mathbf{k}_{\perp}^2l^{2}}{2}\right)
L_{n_<+1}^{|n-n'|}\left(\frac{\mathbf{k}_{\perp}^2l^{2}}{2}\right),
\end{eqnarray}
where $n_<=\text{min}(n,n')$ and $n_>=\text{max}(n,n')$.

\end{document}